\documentclass[aps,twocolumn,groupedaddress,prb]{revtex4}
\usepackage[dvips]{graphicx}
\usepackage[]{amsmath}
\usepackage[]{amssymb}
\usepackage[]{amsfonts}
\usepackage[]{latexsym}
\usepackage[]{float}
\usepackage{bm}

\newcommand{\mib}[1]{\mbox{\boldmath$#1$}}
\newcommand{\bfk}{{\mib k}}

\newcommand{\mat}[4]
{
\left(
\begin{array}{cc}
#1 & #2 \\
#3 & #4 
\end{array}
\right)
}

\newcommand{\mvec}[2]
{
\left(
\begin{array}{c}
#1  \\
#2  
\end{array}
\right)
}

\usepackage[]{endnotes}
\let\footnote=\endnote

\begin{document}


\title{
Topological analysis of the quantum Hall effect in graphene: 
Dirac-Fermi transition across 
van Hove singularities and the edge vs bulk quantum numbers
}


\author{Yasuhiro Hatsugai}
\affiliation{Department of Applied Physics, University of Tokyo, Hongo, 
Tokyo 113-8656, Japan}
\author{Takahiro Fukui}
\affiliation{Department of  Mathematical Sciences, Ibaraki University, Mito 310-8512, Japan}
\author{Hideo Aoki}
\affiliation{Department of Physics, University of Tokyo, Hongo, 
Tokyo 113-0033, Japan}



%
\homepage[The quality of the figures is reduced due to the cond-mat file size limit.
The original pdf file is obtained from the link.
]{http://pothos.t.u-tokyo.ac.jp/Lab/Research/preprints/TopoQHE_honey.pdf
}

\begin{abstract}
Inspired by a recent discovery of a peculiar integer quantum Hall 
effect (QHE) in graphene, we study QHE 
on a honeycomb lattice in terms of the topological quantum number, 
with two-fold interests: 
First, how the zero-mass Dirac QHE around the center of the 
tight-binding band 
crosses over to the ordinary finite-mass fermion QHE 
around the band edges.  
Second, how the bulk QHE is related with the edge QHE
for the entire spectrum including Dirac and ordinary behaviors.  
We find the following: 
(i) The zero-mass Dirac QHE (with $\sigma_{xy}=\mp(2N+1)e^2/h, N$: integer) 
persists, surprisingly, up to the van Hove singularities, at which the 
ordinary fermion behavior abruptly takes over.  
Here a technique developed in the lattice gauge theory enabled us to calculate 
the behavior of the topological number over the entire spectrum.   
This result indicates 
a robustness of the topological quantum number, and should be observable 
if the chemical potential can be varied over a wide range in graphene. 
(ii) To see if the honeycomb lattice is singular in producing the anomalous 
QHE, we have systematically surveyed over square$\leftrightarrow$honeycomb
$\leftrightarrow$$\pi$-flux lattices, which is scanned by 
introducing a diagonal transfer $t'$.  We find that the 
massless Dirac QHE [$propto (2N+1)$] forms a critical line, that is, the
presence of Dirac cones in the Brillouin zone is preserved by the 
inclusion of $t'$ 
and the Dirac 
region sits side by side with ordinary one persists 
all through the transformation.  
(iii) We have compared the bulk QHE number obtained by an adiabatic continuity 
of the Chern number across the square$\leftrightarrow$honeycomb$\leftrightarrow$
$\pi$-flux transformation and numerically obtained edge QHE number 
calculated from the whole energy spectra for sample with edges, 
which shows that the bulk QHE number coincides, as in ordinary 
lattices, with the edge QHE number throughout the lattice transformation.  
\end{abstract}

\maketitle


\section{Introduction}
Electrons on a honeycomb lattice, despite its 
simplicity, provide interesting problems in 
condensed matter physics, especially in its 
topological aspects. 
Field theoretically, Dirac particles and 
associated gauge fields have been intensively investigated from a
topological point of view,\cite{eguchi80,current85} so 
electronic properties for the honeycomb lattice may open 
new avenues for condensed matter phenomena.
Indeed, there have been several proposals about 
peculiar properties in condensed matter systems that have 
zero-mass Dirac particles at low energy scales.\cite{Seme84,Hal87} 
Apart from honeycomb lattice, zero-mass 
Dirac particles appear in condensed matter physics as effective theories 
in various guise. These include the d-wave superconductivity\cite{Lee93,NTW94},
the so-called 
$\pi$-flux or chiral spin state,\cite{Hase89,Wen89wwz}
Anderson localization problems,\cite{Fish84,Lee93} 
spin related problems on the honeycomb lattice,\cite{Ohgushi00,KM05} 
and quantum phase transitions in two dimensions.\cite{Lud93,Hat90} 

A seminal highlight, however, 
is the quantum Hall effect (QHE) in the honeycomb lattice,\cite{ZhengAndo,Gus05}  
which has recently been observed\cite{Nov05,Zha05}  
in graphene, a monolayer graphite with a honeycomb array of carbon 
atoms.   
While the study of graphite has a long history, 
recent advances have been directed toward nanostructured graphite 
such as the carbon nanotube \cite{Ando05} or 
nanographite with boundary magnetism expected 
to arise from edge states.\cite{Fuji96,Waka98}  
In this context, the monolayer graphene is particularly 
interesting as an ideal realization of the honeycomb lattice, 
and the discovery of QHE has kicked off intenseive studies.  

Quite generally, 
topologically nontrivial states are characterized 
not by local order parameters as in symmetry breaking states, 
but by geometrical phases,\cite{Berry84,Simon83} where what are now known as 
topological orders can be realized.\cite{Wen89,Shapere89,Hat04,Hat05}  
One interesting consequence is that topological quantum numbers for the 
bulk can often be related with those for the edge states in finite systems.
With this bulk-edge correspondence 
topological properties which can be hidden in the bulk 
may thus become visible around the edges.  
A typical example is the edge states in QHE 
systems.\cite{Laughlin81,Halperin82,Hatsugai93a,Hatsugai93b}

For the QHE on the honeycomb lattice, 
we can then pose two fundamental questions: 
(i) While the low-energy theory around the band center ($E=0$) is 
that of the zero-mass Dirac particle, which is now realized to 
give an anomalous QHE, how this should be taken over 
by ordinary theories as we go away from the band center?  
(ii) How should the bulk-edge correspondence look like for the 
zero-mass Dirac particle?  The question (ii) 
is of special interest in the context of 
the zero mode of the massless Dirac 
particles. \cite{Wit82,Niem83}  There, a bipartite structure 
(chiral symmetry) in the honeycomb lattice is intimately related to the 
appearance of zero mode edge states.\cite{Ryu02,Ryu03cb} 

As for the question (i), Zheng and Ando\cite{ZhengAndo} 
have numerically calculated 
the QHE on honeycomb with a self-consistent Born approximation.  
They have shown the anomalous QHE around $E=0$, but they have 
also calculated the Hall conductivity for the entire energy region 
in this approximation.  More recently, 
Sheng {\it et al.}\cite{SSW06} have examined the QHE in graphene, computing
the QHE number over the whole energy spectrum. 
They have shown that Dirac-like quantization of the Hall
conductivity appears only near the zero energy and
conventional quantization can be observed in the band edge region.  
However, our question here is on the precise topological quantum 
number\cite{kh} 
in the region including around the boundary between the 
anomalous and ordinary ones.

So in this paper,
we explore the topological aspects for electrons 
on the honeycomb lattice in a magnetic field.  
We show the following: 

(i) The zero-mass Dirac particle behavior (with the Hall conductivity 
$\sigma_{xy}=\pm(2N+1)e^2/h$, where $N$ is an integer and we ignore 
spin degeneracy) persists, 
surprisingly, up to finite energies, at which the usual 
finite-mass fermion behavior abruptly takes over.  The boundary energies are 
identified to be the van Hove singularities.  
Here a technique developed in the lattice gauge theory enabled us to calculate 
the behavior of the topological number, which can become huge 
over the entire spectrum.   
This result indicates 
a robustness of the topological quantum number, and should be observable 
if the chemical potential can be varied over a wide range in graphene. 

(ii) To see if the honeycomb lattice is singular in producing the anomalous 
QHE, we have systematically surveyed the systems that 
extend over square$\leftrightarrow$honeycomb
$\leftrightarrow$$\pi$-flux lattices  by 
introducing a diagonal transfer $t'$.  We find that the 
Dirac region always exists (with its boundary dependent 
on $t'$ and siting side by side with ordinary one) 
all through the transformation (except at the square lattice, 
which, with merged van Hove singularities, 
is rather singular in the present viewpoint).  
It implies the massless Dirac fermion forms a critical line
in a parameter space. It does not occur as a critical point by adjusting
parameters.
Incidentally, we make an interesting observation, in the region 
honeycomb
$\leftrightarrow$$\pi$-flux lattices, that the presence of 
multiple extrema in the band dispersion can give rise to 
a fermion doubling
with the Chern number $\propto 2N$ rather than $(2N+1)$. 

(iii) We have then 
compared (a) the bulk QHE number obtained by an adiabatic continuity 
of the Chern number across the square$\leftrightarrow$honeycomb$\leftrightarrow$
$\pi$-flux transformation and (b) numerically obtained edge QHE number 
calculated from the whole energy spectra for sample with edges.   
The result shows that the bulk QHE number coincides, as in ordinary 
lattices, with the edge QHE number throughout the lattice transformation.  
Incidentally, the $E=0$ flat edge mode persists in strong magnetic 
fields.

The organization of the paper is as follows:  
In the next section, we define the tight-binding model on the honeycomb 
lattice, where we introduce a diagonal 
transfer to tune the position of 
the van Hove singularities to go over to the square and 
$\pi$-flux lattices. 
We present numerical results for the energy spectra as a function of
a magnetic field (Hofstadter diagram), which enables us to infer 
a topological relationship among the honeycomb, square and $\pi$-flux lattices.  
In Sec. \ref{s:Bul}, 
we compute the Hall conductivity of the bulk as a 
function of the chemical potential, based on a 
lattice-gauge theoretical method developed in 
Ref. \cite{Fukui05}. 
This calculation is  
manifestly gauge invariant and guarantees integer Chern numbers, 
so that the method is powerful 
in evaluating the QHE topological number over the whole spectrum, 
including the van Hove singularities, which turn out 
to accompany singular behaviors in QHE.  
There the topological equivalence in Sec. \ref{s:TopEqu} 
is thus confirmed with
respect to the bulk topological properties.
In Sec. \ref{s:DioPha}, we show that 
the conversion square$\leftrightarrow$honeycomb
$\leftrightarrow$$\pi$-flux 
has a virtue of enabling us to derive the Diophantine
equation for the Dirac-fermion regime from the 
adiabatic principle for the topological quantum number, 
which exists as far as the energy gap in which $E_{\rm F}$ reside 
does not close.  
Sec. \ref{s:EdgSta} is devoted to the edge states of the model.
Edge states in finite (cylindrical) systems are analyzed with 
the transfer matrix formalism, and 
we show that edge states, with Laughlin's argument,\cite{Laughlin81}
indeed reveal the Dirac like 
quantization sitting next to the conventional quantization separated 
by van Hove singularities.
Thus, the bulk-edge correspondence is 
confirmed.
A summary and discussion is given in Sec. \ref{s:Sum}.



\section{Model and topological equivalence}\label{s:Ham}
Let us start with defining the Hamiltonian for 
a model that interpolates honeycomb lattice 
with square and $\pi$-flux lattices.  
This parameter plays a key role in examining the topological
properties in terms of the adiabatic principle.

\subsection{Model}
\begin{figure}[htb]
\begin{center}
\includegraphics[width=0.6\linewidth]{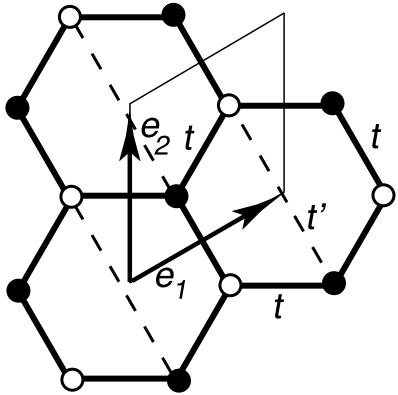}
\end{center}
\caption{A honeycomb lattice with a unit cell 
and a extra transfer $t'$ indicated. 
} 
\label{f:unit}
\end{figure}

The Hamiltonian for two-dimensional tight-binding systems in 
a uniform magnetic field $\mib{B}$ applied normal to the plane is
\begin{eqnarray*}
H=H_{\rm honeycomb}+H'.
\end{eqnarray*}
Here, $H_{\rm honeycomb}$ is the tight-binding model on the honeycomb lattice
with nearest-neighbor hopping,
\begin{eqnarray*}
H_{\rm honeycomb} &=t&\sum_{j} \Big[
c ^\dagger _\bullet(j)c_\circ(j)
+e^{i2\pi \phi j_1}c ^\dagger _\bullet(j)c_\circ(j-e_2)
\nonumber\\
&&
+c ^\dagger _\bullet(j+ e_1)c_\circ(j)
\Big] +\mbox{h.c.},
\end{eqnarray*}
where $t=-1$ is the transfer (taken to be the unit of energy).  
Since the honeycomb, a non-Bravais lattice with two sites per unit cell 
is bipartite, we have defined two kinds of fermion operators, 
$c_\circ(j)$ and $c_\bullet (j)$ as in Fig. \ref{f:unit}, 
where $j= j_1e_1+j_2e_2$ labels the unit cell 
with two translation vectors $e_1 = (3/2 ,\sqrt{3}/2)a$ and
$e_2 = (0,{\sqrt{3}} )a$.  
The magnetic field is characterized by
\begin{equation*}
\phi = BS_6/(2\pi) \equiv p/q,
\end{equation*}
the magnetic flux (assumed to be rational) penetrating each hexagon of 
area $S_6=(3\sqrt{3}/2)a^2$.  
we have adopted the Landau gauge for the vector potential.  
For simplicity we neglect the spin degrees of freedom, since the 
graphene has a very small Zeeman splitting.
\begin{figure*}[htb]
\begin{center}
\includegraphics[width=0.99\linewidth]{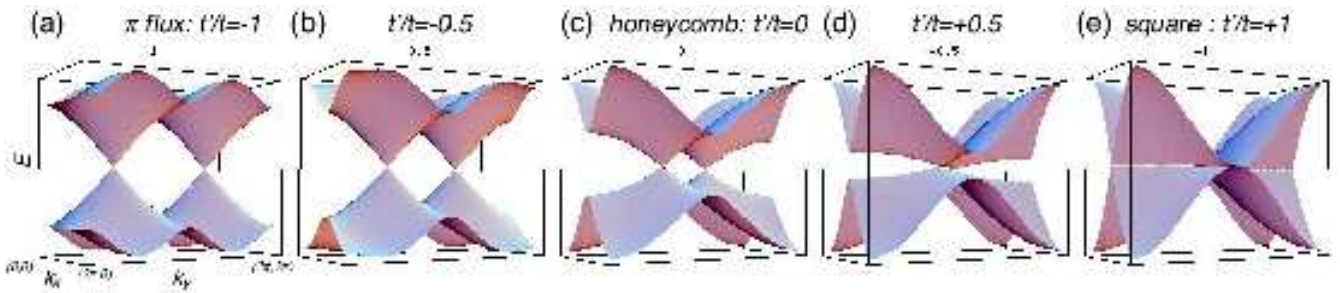}
\end{center}
\caption{
Energy dispersions
for 
(a) $t'/t=-1$ ($\pi$-flux lattice),
(b) $t'/t=0.5$, 
(c) $t'/t=0$ (honeycomb), 
(d) $t'/t=0.5$, and 
(e) $t'/t=1$ (square).
}
\label{f:disp}
\end{figure*}
\begin{figure*}[htb]
\begin{center}
\includegraphics[width=0.99\linewidth]{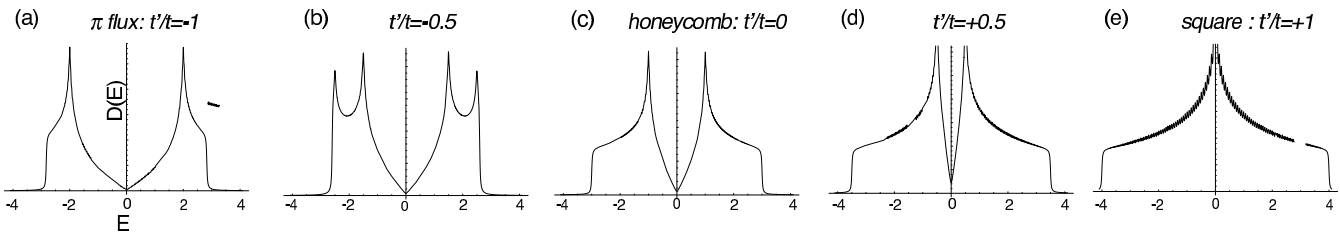}
\end{center}
\caption{Density of states for 
(a) $t'/t=-1$ ($\pi$-flux lattice),
(b) $t'/t=0.5$, 
(c) $t'/t=0$ (honeycomb), 
(d) $t'/t=0.5$, and 
(e) $t'/t=1$ (square).
}
\label{f:dos}
\end{figure*}
If we add a hopping across each hexagon,
\begin{eqnarray*}
H'=t'\sum_je^{-i2\pi \phi (j_1+1/2)}c^\dagger_\bullet(j+e_1-e_2)c_\circ(j)
+\mbox{h.c.},
\end{eqnarray*}
we can change the system continuously from $t'/t= -1$ 
(which is referred to as $\pi$-flux lattice) 
to $0$ (honeycomb) and $1$ (square).

\begin{figure*}[htb]
\begin{center}
\includegraphics[width=0.99\linewidth]{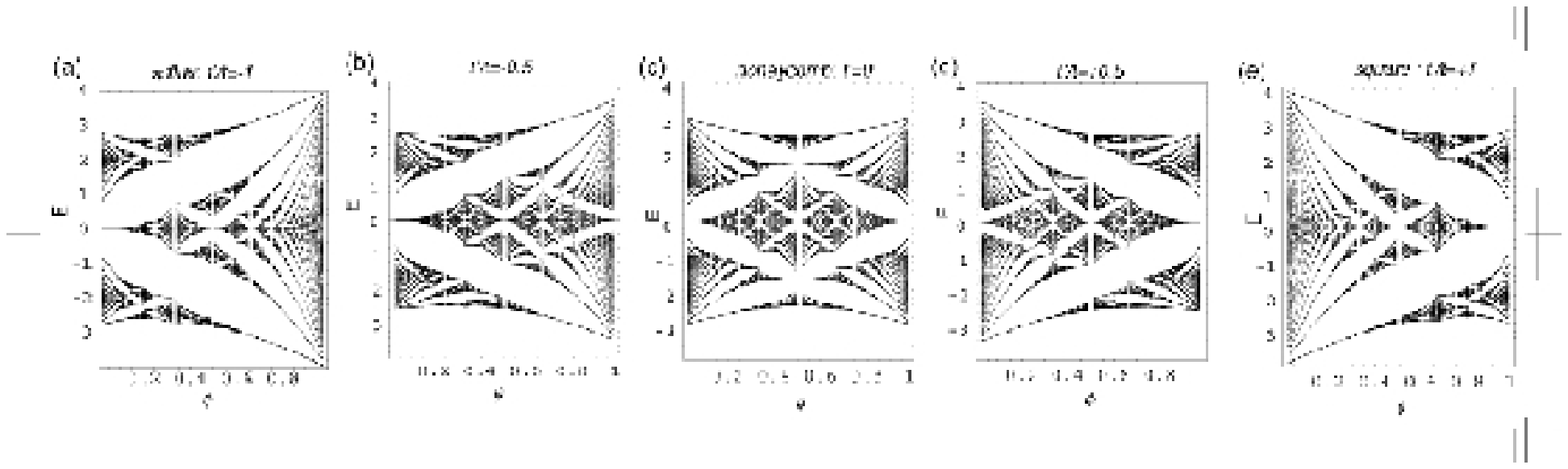}
\end{center}
\caption{Hofstadter's diagram (energy spectrum against the magnetic 
flux $\phi$) for 
(a) $t'/t=-1$ ($\pi$-flux lattice),
(b) $t'/t=0.5$, 
(c) $t'/t=0$ (honeycomb), 
(d) $t'/t=0.5$, and 
(e) $t'/t=1$ (square).
}
\label{f:hof}
\end{figure*}

In the momentum space, the Hamiltonian for $\phi=p/q$ is expressed as
\begin{eqnarray}
H=
\int_0^{2\pi/q}\frac{dk_1}{{2\pi/q}}
\int_0^{2\pi}  \frac{dk_2}{{2\pi}}~
\bm{c}^\dagger(\bfk)h(\bfk)\bm{c}(\bfk),
\label{HamMom}
\end{eqnarray}
where $\bm{c}^\dagger(\bfk)$ is a $2q$ dimenisional vector and 
$h(\bfk)$ is a $2q\times 2q$ matrix (see Appendix \ref{s:AppMom}).  
In zero magnetic field, we have
\begin{alignat*}{1} 
h(\bfk)=& 
t\mat{0}{\Delta(\bfk)}{\Delta^*(\bfk)}{0},
\\
\Delta(\bfk)=& 1 + e^{ik_2}+e^{ik_1} \left[1 +(t'/t) e^{-i k_2}\right] .
\end{alignat*}
To see how $t'$ shifts the position of 
van Hove singularities, we show the energy dispersions and
the density of states in the zero magnetic field in Figs. \ref{f:disp} and \ref{f:dos}.  
The square lattice has a van Hove singularity at the band center, 
which splits into two as we decrease $t'/t$ from 1. 
As we shall stress, it is between the two singularities that 
a zero-mass Dirac like gapless dispersion appears in the honeycomb lattice.  
We can in fact rigorously show \cite{commentcircle} that 
the zero gap with a linearly vanishing density of states 
around $E=0$ is not an accident for honeycomb, 
but exists for $-3 < t'< 1$, so such a situation 
persists down to the $\pi$-flux lattice.

Ordinary band dispersions and density of states reside 
outside the van Hove singularities. 
So our question here is how the quantum Hall effect 
should look like along the sequence, 
square$\leftrightarrow$honeycomb $\leftrightarrow$$\pi$-flux.

\begin{figure*}[hbt]
\begin{center}
\includegraphics[width=0.99\linewidth]{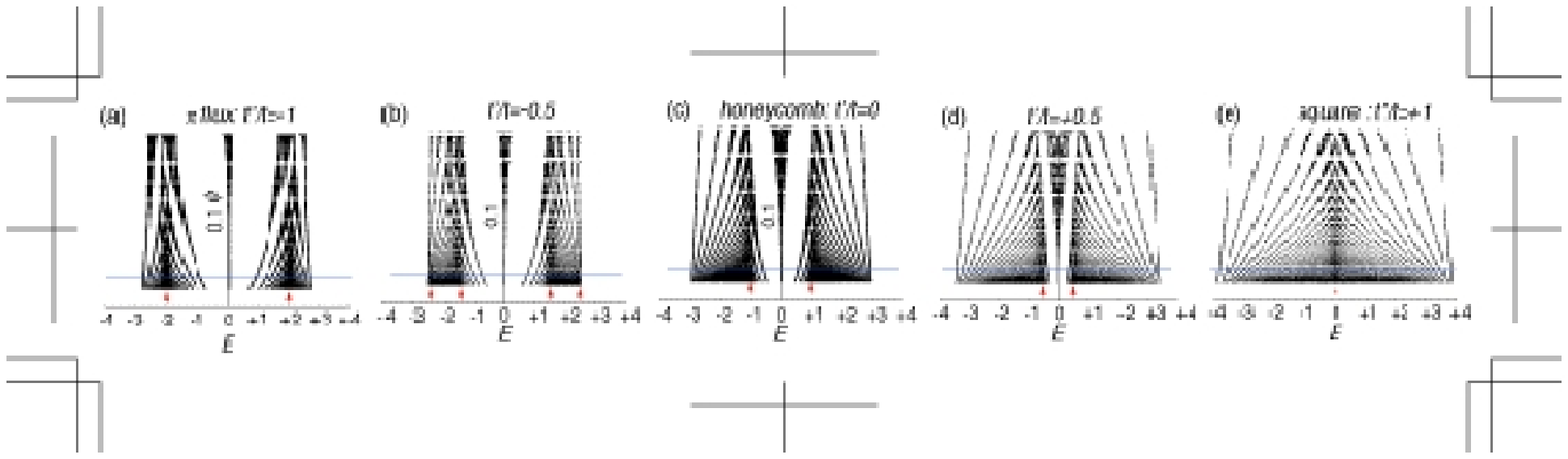} 
\end{center}
\caption{Blowup of Hofstadter's diagram in a weak magnetic 
field region for 
(a) $t'/t=-1$ ($\pi$-flux), 
(b) $t'/t=0.5$, 
(c) $t'/t=0$ (honeycomb), 
(d) $t'/t=0.5$ and $t'/t=1$
(square). 
The arrows indicate the positions of van-Hove singularities.
The blue lines indicate  positions of a flux $\phi=1/31 $, which corresponds to
the one in the Figs.\ref{f:chern}.
}
\label{f:ToPi}
\end{figure*}

\begin{figure*}[htb]
\begin{center}
\includegraphics[width=0.99\linewidth]{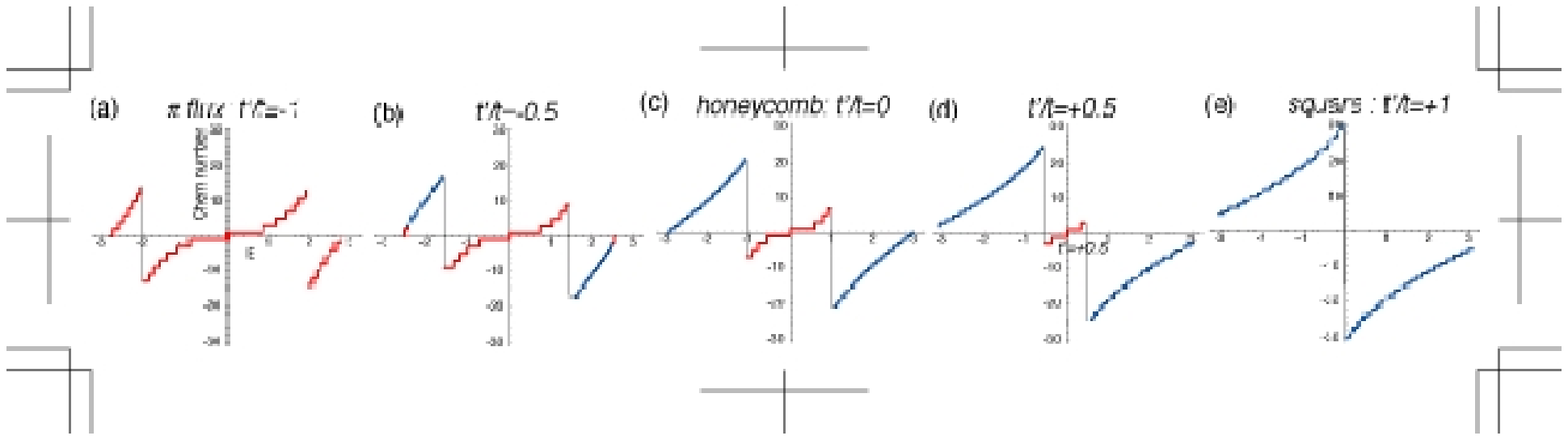} 
\end{center}
\caption{ The Chern number (Hall conductivity in unit of $-e^2/h$) 
for magnetic field $\phi=1/31$ is plotted against 
the Fermi energy $E_{\rm F}$: for 
(a) $t'/t=-1$ ($\pi$-flux),
(b) $t'/t=0.5$, 
(c) $t'/t=0$ (honeycomb), 
(d) $t'/t=0.5$, and 
(e) $t'/t=1$ (square).
The red lines indicate the steps of two in the Chern number sequence 
($\sigma_{xy}=\pm(2N+1)e^2/h, N$: integer), 
while the blue lines step of one ($\sigma_{xy}=\pm Ne^2/h$).
}
\label{f:chern}
\end{figure*}

\subsection{Topological Equivalences} \label{s:TopEqu}
In a magnetic field $B$, the spectra of the 
lattice Hamiltonians against $B\propto \phi$ 
take fractal shapes, usually called 
Hofstadter's diagram,\cite{Hof76} where hierarchical series of 
energy gaps exist.  The butterfly is deformed as $t'$ is varied.  
Here, an adiabatic principle plays a crucial role: 
it dictates that one can keep track of quantum mechanical ground states 
when the Hamiltonian is transformed with a continuous change 
of parameter(s) {\it if there is a gap above the ground state and 
if the gap remains finite throughout}.\cite{Hat90} 
Let us apply this argument to the present model, where 
$t'$ is the adiabatic parameter.  
In Hofstadter's butterfly we can see 
many Landau bands. The change of $t'$ makes 
some of the bands merge (or, more precisely, some of the 
gaps between Landau bands merge, since the 
spectrum is fractal). 
The Landau levels are characterized by the quantum Hall numbers, 
which are topological (Chern) numbers as 
will be discussed in Sec. \ref{s:Bul}, 
and the numbers remain unchanged against the adiabatic change 
as long as the gap in which $E_{\rm F}$ reside does not collapse.  
This is a topological stability.

\subsubsection{Topological equivalences between van-Hove
singularities}

The Hofstadter diagram for honeycomb lattice was first obtained in  
Ref. \cite{ram}.  Rammal 
has already noted the presence of the $E=0$ Landau level 
which is outside Onsager's semiclassical quantization scheme.  
The spectrum for the honeycomb lattice is symmetric about $\phi=\pi$.  
In Figs. \ref{f:hof}, we show how the spectrum 
versus $\phi\propto B$ is deformed with  $t'$ 
for the square lattice ($t'/t\rightarrow 1$), 
or for the $\pi$-flux lattice  
($t'/t\rightarrow -1$).  
In the present construction the flux per plaquette on the square 
lattice corresponds to half the hexagon in the honeycomb lattice, 
which is why Hofstadter's diagram has a period of $4\pi$ for $t'\neq 0$. 
The property that the spectrum is invariant against 
$\phi\to-\phi$ as well as against 
$t' \leftrightarrow  -t', \phi \leftrightarrow 1-\phi$
is retained throughout.  
The honeycomb lattice with $t'=0$ corresponds to the self-dual point 
for the $t' \leftrightarrow  -t'$ symmetry.  

We can immediately notice from Fig. \ref{f:hof}(a)(b) that the large gaps 
just above and below the zero energy remain. 
This implies that the topological number should remain the same 
when $E_{\rm F}$ lies in the gap.  
A closer examination shows, surprisingly, that 
this holds for other gaps, all the way up to a finite energy, $E_c$, 
in fact.  
Figure \ref{f:ToPi} is a blowup of the low magnetic field region.  
We can see that there is no level crossing for other energy gaps as well 
in the energy region $E\leq E_c$, so that 
we can keep track of the main gaps in the adiabatic change 
between the honeycomb and $\pi$-flux ($t'/t = 0 \leftrightarrow -1$).  
The boundary energy $E_c$ is seen to reside around the region 
where the Landau level fan extending from $E=0$ 
meets another Landau fan extending from the band 
edges $|E| \simeq 3$.  
We shall show below, by using the calculation of the quantum Hall numbers, 
that we can identify $E_c$  as the van Hove energies (which are functions 
of $t')$.

\subsubsection{Topological equivalences near band edges}

The above situation is exactly symmetric for the 
crossover of the honeycomb $\leftrightarrow$ square 
($t'/t=0\leftrightarrow1$),
where we have only to replace $\phi$ with $1 - \phi$. 
This means that, in the weak magnetic field region 
depicted in Fig. \ref{f:ToPi}, 
large gaps around $E=0$ are closed as we go to the square lattice, 
while we can keep track of the 
Landau fan starting from the band edge.
This property is in fact highly nontrivial, since the 
Landau levels is usually analyzed in the effective-mass 
approach only near the band center or band edges.  
This is exactly why we want to confirm the topological equivalences 
in terms of the quantum Hall number in the next section.
 


\section{Topological Equivalences in the Bulk} \label{s:Bul}

\subsection{Hall condactance}

It is now firmly established that the Hall conductivity of the noninteracting 
2D electron systems, as described with the 
linear response theory,\cite{Aoki81} 
may be regarded as a topological quantum 
number.\cite{Thouless82,Kohmoto85}  
Namely, when the Fermi energy lies in 
the $n$-th gap, the Hall conductivity is given by 
\begin{eqnarray*}
\sigma_{xy}=-\frac{e^2}{h} c_{\rm F}(E_{\rm F}),
\end{eqnarray*}
where $c_{{\rm F}}$ is an integer called Chern number, 
which describes how the wave function (a vector bundle) 
responds to a vector potential 
generated by Berry's gauge potential in the Brillouin zone.  
To compute the number, we need to diagonalize the Hamiltonian 
\begin{eqnarray*}
h(\bfk)\mib{\varphi}_j(\bfk)=\epsilon_j(\bfk)\mib{\varphi}_j(\bfk),\quad (j=1,2,\cdots, 2q),
\end{eqnarray*}
where $h(\bfk)$ is the $2q\times 2q$ matrix defined in Eq.(\ref{HamEle}), 
$\mib{\varphi}_j(\bfk)$ an eigenvector, and we assume that 
$\epsilon_1(k)\le\cdots
\le\epsilon_n(k)<E_{\rm F}<\epsilon_{n+1}(k)\le\cdots\le\epsilon_{2q}(k)$
holds over the entire Brillouin zone.  
We can then define Berry's gauge potential, 
\begin{eqnarray}
\mib{A}_{ij}(\bfk)=\mib{\varphi}^\dagger_i (\bfk)\nabla_{\bfk} 
\mib{\varphi}_j(\bfk) , \quad 
(1\le i,j\le n),
\label{ConGauPot}
\end{eqnarray}
where $\mib{A}_{ij}(\bfk)$ is an anti-Hermitian $n\times n$ matrix.
The Chern number is given as 
the U(1) part of the above U($n$) gauge potential,
\begin{eqnarray}
c_{{\rm F}}(E_{\rm F})=\frac{1}{2\pi i}\int{\rm Tr}~d A ,
\label{ConCheNum}
\end{eqnarray}
where $A(\bfk)\equiv A_\mu(\bfk)dk_\mu$ is a one-form .

This formulation, a non-Abelian 
extension of Berry's gauge potential,\cite{Hat04,Hat05} 
may seem too elaborate, 
but is useful when there are multiple bands below the Fermi energy.   
Namely, this formula holds {\it even if} 
some gaps in the Fermi sea 
are closed, as long as the bands in question, 
$\epsilon_n(\bfk)$ and $\epsilon_{n+1}(\bfk)$, do not cross.  
In the special case when all the bands in the Fermi sea are 
separated with each other, 
the Chern number, Eq.(\ref{ConCheNum}), is simply the sum of 
the Chern numbers assigned to individual bands,\cite{Kohmoto85} 
but we have opted for the above formula, since some of 
the gaps do collapse as $t'$ is varied, as we shall see.

\subsection{Lattice gauge theory technique}

To actually evaluate the topological integer for the present system, we need to 
calculate eigenfunctions of the $2q\times 2q$ Hamiltonian (\ref{HamEle}).
Diagonalization can be done only numerically, so that 
the eigenfunctions are obtained in practice on mesh 
points in the Brillouin zone.  This can cause a serious problem 
if we want to obtain the Chern numbers, especially around the van Hove 
singularities, since they can behave wildly there.  
The problem becomes even more formidable if we want to maintain 
the calculation manifestly gauge invariant and pin point the 
Chern numbers as integers.  
Here we adopt a method,\cite{Fukui05,Hat05ep2ds} developed recently 
in the context of the {\it lattice gauge theory}. 
\cite{Lus82,Phi85,PhiSto86,PhiSto90,Lus99,FSW01} 
This implementation precisely guarantees manifest gauge invariance 
and integer Chern numbers.

Let us first compute the eigenfunction of the Hamiltonian on 
meshes in the Brillouin zone.  
In what follows, we choose the Landau gauge for the magnetic field,
so that the mesh points are denoted as $(k_{j_1},k_{j_2})$, where 
discrete sets of momenta with $j_\mu=0,\ldots,N_\mu-1$ are defined 
by $k_{j_1}=2\pi j_1/(qN_1)$ and $k_{j_2}=2\pi j_2/N_2$ 
in the Brillouin zone extending over $0\le k_1<2\pi/q$ and $0\le k_2<2\pi$.
Diagonalization of $h(\bfk)$ on these sites 
$\bfk=\bfk_\ell\equiv (k_{j_1},k_{j_2}), \ell=1, \dots, N_1N_2$ 
provides eigenfunctions with 
$
h(\bfk_\ell)\mib{\varphi}_j(\bfk_\ell)=
\epsilon_j(\bfk_\ell)\mib{\varphi}_j(\bfk_\ell)
$, 
which can be chosen
to satisfy the periodic boundary condition 
$\mib{\varphi}_j(\bfk_\ell+N_\mu\hat{\mib{\mu}})=\mib{\varphi}_j(\bfk_\ell)$,
where $\hat{\mib{\mu}}$ is a vector of length 
$|\hat{\mib{1}}|=2\pi/(qN_1)$ and $|\hat{\mib{2}}|=2\pi/N_2$ 
along $\bfk_\mu$.

Provided that the Fermi energy lies between the $n$th gap 
(namely, $\epsilon_n(\bfk_\ell)<E_{\rm F}<\epsilon_{n+1}(\bfk_\ell)$ 
for every $\bfk_\ell$), and hence the Fermi sea is composed of $n$ bands 
(some of which may merge), 
we define a $U(1)$ link variable of the Fermi sea as
\begin{eqnarray*} 
U_\mu(\bfk_\ell) 
&\equiv& |\det\bm{U}_\mu(\bfk_\ell)|^{-1}
\det\nolimits {\bm U_\mu(\bfk_\ell)} ,
\end{eqnarray*}
where
\begin{eqnarray*} 
({\bm U}_\mu)_{ij}
&=&
\mib{\varphi}_i^\dagger(k_\ell)\mib{\varphi}_j
(\bfk_\ell+\hat{\mib{\mu}}),
\quad(1\le i,j\le n ) .
\end{eqnarray*}
The link variables are well-defined except at 
singular points 
$\det\bm{U}_\mu(\bfk_\ell) = 0$, which corresponding to 
``vortices" and ``antivortices" in the wave function, 
and we can always make the mesh avoid them with an 
infinitesimal shift. 
With the link variable we can define a lattice field strength by
\begin{alignat*}{1} 
F_{12}(\bfk_\ell)
\equiv &
\ln U_1(\bfk_\ell)U_2(\bfk_\ell+\hat{\mib{1}})
U_1(\bfk_\ell+\hat{\mib{2}})^{-1}U_2(\bfk_\ell)^{-1},
\end{alignat*} 
where the principal branch of the
logarithm with $-\pi<{F}_{12}(\bfk_\ell)/i\leq\pi$ is taken.  
The field strength is by definition invariant under gauge
transformations, so that no specific gauge fixing is required.  
The Chern number on the lattice is defined as 
\begin{alignat}{1} 
c_{\rm F}(E_{\rm F}) = 
& \frac{1}{2\pi i}\sum_\ell F_{12}(\bfk_\ell). 
\label{LatCheNum}
\end{alignat} 
The Chern number thus computed is strictly an integer, 
since $\sum_\ell F_{12}$ just counts the number of vortices 
minus the number of antivortices in the 
Brillouin zone (i.e., the number of times the branches in ln are 
traversed), as has been proved in Ref. \cite{Fukui05}.
Therefore, it should give the number in the continuum, $c_{\rm F}$ 
appearing in Eq. (\ref{ConCheNum}) 
when the mesh in the Brillouin zone is sufficiently dense.  
This is due to the integral character of the lattice Chern number: 
For large $N_\mu$, we have
\begin{alignat*}{1}
U_\mu(\bfk)=1+{\rm Tr}A_\mu(\bfk_\ell) \delta k_\mu,
\end{alignat*}
with $A_\mu$ defined by Eq.(\ref{ConGauPot}), and we have
\begin{alignat*}{1} 
F_{12} =&
\ln \prod_{\square} 
(1+{\rm Tr} A_\mu\delta k_\mu)+{\cal O}(|\delta k|^3 )
\nonumber\\
=&{\rm Tr }\left[\partial_1A_2(k)-\partial_2A_1(k)\right]
\delta k_1 \delta k_2+{\cal O}(|\delta k|^3 ) .
\end{alignat*} 
When coupled with Eq. (\ref{LatCheNum}), this reduces to
Eq. (\ref{ConCheNum}), the Chern number in the continuum Brillouin zone.

\subsection{Dirac vs ordinary fermion quantization}
\label{s:CheNumDvsF}

Let us now show the numerical results 
for the Chern number $c_{\rm F}$ in Eq. (\ref{LatCheNum})
as a function of the chemical potential $E_{\rm F}$ for 
various values of $t'$. 
While the integral Chern number $c_{\rm F}$ is defined 
only for each gap, we have plotted $c_{\rm F}$ as step functions, 
which makes sense as long as 
the magnetic field is not too large, as in Fig. \ref{f:chern} 
with $\phi=1/31$, since the width of each Landau band 
is then much smaller than the size of gaps.  
We have displayed the values for spinless fermions: 

The result shows a striking answer to 
one of our key questions: what is the fate of the 
Dirac-like behavior as we go away from $E=0$.  
When the Fermi energy is swept in the honeycomb lattice, 
the Dirac-like Hall conductivity steps of two
(or four when spin degeneracy is included) 
in units of $-e^2/h$ exist around $E=0$ as has been noted by many 
authors.\cite{Gus05,Nov05,Zha05,SSW06}  Let us call this Dirac-Landau 
behavior.  
Now, we can see that this Dirac-Landau behavior persists 
{\it all the way up to} some energy, which we identify to be 
the van Hove singularities appearing in Fig. \ref{f:dos}.  
At these energies we have then a {\it huge step} accompanied by a 
sign change in the Hall conductivity.  
This result implies the following: 
The effective theory near the zero energy
is Dirac-like fermions in the continuum limit, for which 
an unconventional quantization of the Hall conductivity has been derived 
from the Dirac Landau levels.\cite{ZhengAndo,Gus05}  
The present calculation leads to the conclusion that the unusual 
property extends to an unexpectedly wide region of energy 
in the actual lattice fermion system.
 Outside the van Hove energies (i.e., in the band-edge regions), 
we recover the conventional QHE where the step changes 
by one in units of $-e^2/h$ (which we now call Fermi-Landau).  
This implies that a huge step accompanied by a 
sign change has to occur at the boundary 
between Dirac-Landau and Fermi-Landau 
behaviors, 
or the bands just at the van-Hove singularities 
induce a change, which is topological in that the 
relevant quantum numbers are topological. 

A second striking feature in Fig. \ref{f:chern} is that 
the Dirac-like Hall conductivity $\propto(2N+1)$ 
persists all along  $-1\le t'/t<1$ in the $\pi$-flux$\leftrightarrow$
honeycomb$\leftrightarrow$square sequence, except at 
the square lattice.   Namely, 
there always exists the Dirac-Landau behavior 
between the (innermost) van Hove singularities, except 
for the case when the singularities merge at $t'=0$.  
We can check from the band dispersion (Fig.\ref{f:disp}) that 
in this whole region we have two zero-mass Dirac cones 
in the Brillouin zone, which should cause this persistence 
of the anomalous QHE number.


We can now summarize the Hall conductivity for the spinless fermions on 
the honeycomb lattice for the entire energy region as
\begin{widetext}
\begin{alignat*}{1} 
\sigma_{xy}^{\rm hc} =& -\frac {e^2}{h} \times
\left\{
\begin{array}{lll}
+(N+1)    & 
E_{\rm F}<-|t|,  &N =0,1,2,\cdots: 
{\text{ Landau level index counted from the band bottom}}
\\
-(2N+1)    & 
-|t|<E_{\rm F}<0, 
& N =0,1,2,\cdots:
 { \text{ Dirac-Landau level index counted from the zero energy }}
\\
+(2N+1)    & 
0<E_{\rm F}<+|t|, 
& N =0,1,2,\cdots:
 { \text{ Dirac-Landau level index counted from the zero energy }}
\\
-(N+1)    & 
|t|<E_{\rm F},
&  N =0,1,2,\cdots: {\text{ Landau level index counted from the band top}}
\end{array}
\right.
\end{alignat*} 
\end{widetext} 
We should again double these numbers if we include the spin degeneracy.
While the unconventional quantization around the zero energy has been
beautifully observed experimentally \cite{Nov05,Zha05}, 
pushing the chemical potential further to approach the van Hove energies 
should detect the topological phase transition.

We can note that, to be precise, there are in fact 
two sets of van Hove singularities (i.e., four in total, 
see Fig.\ref{f:chern}(b)) 
in the honeycomb$\rightarrow \pi$-flux region.  
There, we have the Dirac-Landau behavior in the region 
between the innermost singularities,
 Fermi-Landau behavior 
in the next region, and another 
behavior 
outside the outermost singularities
with the doubled Chern number $\propto 2N$.
Although this may seem a 
trivial detail, we can in fact make an interesting observation. 
Namely, we can raise a question: how the band-edge region can have 
doubled Chern number than Fermi-Landau behavior?  
We can identify the origin of the doubled Chern numbers 
in the fermion band as the presence of 
two minima (maxima) in the dispersion in the Brillouin zone 
at the band bottom (top).  
So we can make an observation that {\it a degenerate minima (or maxima) 
in the band dispersion can give rise to a  
Fermion species multiplication}.  
We shall reinforce this view 
in terms of the topological equivalences in Sec. \ref{s:TopEqu} 
and edge state pictures in Sec. \ref{s:EdgSta}.

\subsection{Degeneracy of Landau levels and the lattice effect}

Let us now look more closely at the 
distinction between the Dirac-Landau and Fermi-Landau 
levels in terms of the degeneracy of the Landau bands. 
The physical reason why we have a step of two in the 
quantized Hall conductivity for the Dirac-Landau
levels can be ascribed, as in many other literatures, 
to the degeneracy (fermion doubling) of the bands.  
Let us take an example of $\phi=1/31$ employed in
Fig. \ref{f:chern}(c), for which we have 
$2q=62$ eigenvalues $\epsilon_j(k)$. 
Among them, there are 7 pairs of twofold 
degenerate levels around $E=0$ with a total Chern number of 2, while other 48 levels
are nondegenerate. Two of the nondegenerate Landau levels 
are located at the van Hove energies and carry 
a huge Chern number of $-30$. It is such large numbers that 
require the accurate estimate with the present, 
lattice-gauge numerical algorithm.  

The level adjacent to these have the ordinary Chern number of 1 each.  
So this abrupt change in the topological number 
defines the clear boundary.  
 To be precise, however, we can note that 
approximate degeneracy in each two-fold Dirac-Landau levels 
(which is exact for a Dirac fermion with a rigorously linear dispersion) 
become slightly lifted toward the van Hove energy, 
while the gap across the next two-fold level becomes smaller, 
where the splitting ($0.05t$ for $\phi=1/31$)and the gap ($0.02t$) 
become comparable (which is still much smaller than 
the thickness of the line that plots the energy dependence 
of the Chern number in Fig.\ref{f:chern}, so the figure still makes sense).  
The deviation 
 becomes more pronounced for 
stronger magnetic fields.
We shall discuss this feature in Sec. \ref{s:EdgSta} (see, e.g., 
Fig. \ref{f:edges_honey_full} (a) for $\phi=1/5$), where 
the gap can become comparable with the Landau band width.


\section{Diophantine Equation}\label{s:DioPha}

Having numerically calculated the Chern numbers, we can now 
raise the question: can we calculate these numbers 
algebraically?  
For square lattice systems, 
the Diophantine equation according to Thouless {\it et al.}\cite{Thouless82} 
is known to be a simple yet powerful tool to compute the quantum 
Hall number of the Fermi seas.  
For the honeycomb lattice, however, the Diophantine equation 
has not been obtained.  This is exactly where we can exploit 
the topological equivalence establised in the previous sections 
to calculate the
Chern numbers by the use of the Diophantine equation.  
Natural interest here is how the 
Dirac-Landau quantum Hall number of $(2N+1)$ comes about.  
So we start with 
the Diophantine equation for the square lattice, where 
the Chern number $c_J$
of the Fermi sea composed of $J$ bands with flux per plaquette 
$\Phi=P/Q$ is given by the formula
\begin{alignat}{1} 
J \equiv & P c_J\ \ \text{(mod } Q),\ \ |c_J|\le Q/2.
\label{DioPha}
\end{alignat}


Now let us make its adiabatic continuation to the honeycomb lattice with
magnetic field $\phi=1/q$.  
Since the $\pi$-flux lattice is constructed here from 
the honeycomb lattice
by the introduction of $t'/t=-1$, the flux per square plaquette 
corresponds to $\phi/2$ for the hexagonal plaquette 
with the shift in the period discussed in Sec. \ref{s:TopEqu}. 
Therefore, total flux per square plaquette
of the $\pi$-flux lattice,  $\Phi$, reads 
\begin{alignat*}{1} 
\Phi = \frac{P}{Q} =& \frac {1}{2}+ \frac  \phi 2
= \frac {q+1}{2q},
\end{alignat*} 
where the $1/2$ flux in the above equation is due to the $\pi$-flux.
Let us first take the case of $q+1$: prime ($q$: even). Then we have
\begin{alignat}{1} 
P =& q+1,
\ \
Q = 2q.
\label{DioPhaFlu}
\end{alignat} 
Equation (\ref{DioPha}) together with Eq. (\ref{DioPhaFlu}) is 
a Diophantine equation for the honeycomb lattice with even $q$
valid in the Dirac fermion regime.
To identify $J$ we assume for the honeycomb lattice that 
there exist a zero energy Dirac-Landau level, 
and the Dirac-Landau levels between the van Hove singularities
are twofold degenerate. 

Now let us calculate the Hall conductivity around the zero energy 
for the honeycomb lattice 
when $E_{\rm F}$ lies in the $(N+1)$-th gap from the center (e.g., 
$N=0$ is the first gap above the zero energy).  
Note that 
the total number of the bands is $2q$, so that
the number of the negative-energy bands is $q-1$. 
Hence 
the number of bands below the $(N+1)$-th gap is $q-1+2(N+1)$, i.e, 
$J = q +2N+1$.  
From Eq. (\ref{DioPha}), the Diophantine equation then reads 
\begin{alignat*}{1} 
q + 2N +1 \equiv &  (q+1)c_{\rm F}, \quad \text{(mod } 2q),
\end{alignat*} 
or 
\begin{alignat*}{1} 
 (q+1)(c_{\rm F}-1) \equiv & 2N\equiv 2N(q+1), \quad \text{(mod } 2q).
\end{alignat*} 
This has an obvious solution, 
\begin{alignat*}{1} 
c_{\rm F} =&  2N+1:\ E_F>0,
\end{alignat*} 
and its particle-hole transformed $c_{\rm F} = -(2N+1)$ 
for $E_{\rm F}<0$.  
Remarkably, this precisely gives the Chern number of the Dirac fermion, 
as computed in the previous section. 

For an odd $q$, we put
\begin{alignat*}{1} 
q =& 2n +1,\ n=1,2,\cdots.
\end{alignat*} 
Then the mapping to the $\pi$-flux lattice reads
\begin{alignat*}{1} 
\Phi =\frac {P}{Q} =& \frac {q+1}{2q}  
= \frac {n+1}{2n+1}.
\end{alignat*} 
Namely, we have
\begin{alignat}{1}
P=n+1, \quad Q=2n+1.
\label{DioPhaFluOdd}
\end{alignat}
Note that the total number of the bands in this case 
is $Q=q=2n+1$ rather than $2q$ 
in the adiabatically transformed $\pi$ flux system.
Then, the number of bands in the negative energy is $n$,  
and the number of bands below the $(N+1)$-th gap is $n+(N+1)$, i.e,
\begin{alignat*}{1} 
J =& n+1+N,
\end{alignat*} 
which gives, via Eqs.(\ref{DioPha}),(\ref{DioPhaFluOdd}),
a Diophantine equation,
\begin{alignat*}{1} 
n +1+N \equiv & (n+1)c_{\rm F}, \quad \text{(mod }2n+1),
\end{alignat*} 
or $(n+1)(c_{\rm F}-1) \equiv N$. 
This expression is equivalent to 
$-n (c_{\rm F}-1) \equiv  -2n N$. 
This has a solution $c_{\rm F}= 2 N+1$, 
which gives the same Chern number for the prime $q+1$ case.  
%


\section{Edge vs bulk for the Hall conductivity 
in the honeycomb lattice} \label{s:EdgSta}

When a system has a nontrivial topological structure, 
they should generically exhibit 
characteristic properties related to edge states. 
For the honeycomb lattice, 
edge states and their flat dispersion 
have been intensively discussed with 
its relevance to spin alignment\cite{Fuji96,Waka98}. 
Another way of saying is that the honeycomb system has a bipartite 
symmetry, 
which guarantees the presence of the zero mode edge states\cite{Ryu02}, i.e., 
there exist macroscopic edge states at the zero energy 
unless the bipartite symmetry is broken.
This gives rise to dispersionless edge states.  
The flat bands are unstable against
 bipartite symmetry breaking (Peierls instability), 
 which is, e.g., realized when an antiferromagnetic spin ordering occurs.  
This may be viewed as a topological origin of the boundary spin moments.\cite{Ryu03cb}

Now, the problem of edge states vs bulk states is particularly 
interesting for the QHE, since the problem of 
how the Dirac-Landau QHE $\propto (2N+1)$ is related to the edge 
states is fundamentally interesting.  
One of the present authors has shown, for 2D 
square (or anisotropic square) systems, that the edge
states whose energy dispersions run across 
the Landau bands have topological QHE numbers, for which 
\[
\sigma_{xy}^{\rm edge} =\sigma_{xy}^{\rm bulk}
\]
by identifying the connection between the topological integers
for the bulk and for the edge states.\cite{Hatsugai93b,yh-qhe-review}
So let us now focus on the edge states of the honeycomb lattice 
in terms of the topological structure.
It also allows us to clarify the bulk-edge correspondence.

\subsection{Transfer matrix formalism}

\begin{figure}[htb]
\begin{center}
\includegraphics[width=0.99\linewidth]{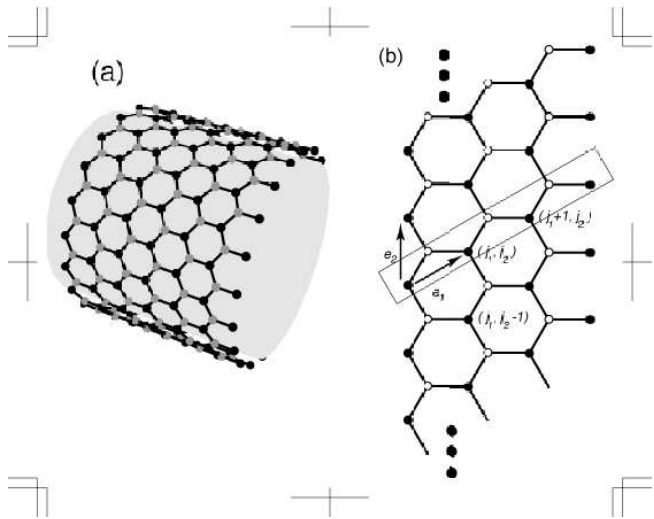}
\end{center}
\caption{A cylindrical system with zigzag and bearded edges.
}
\label{f:zigzag_with_eq}
\end{figure}

\subsubsection{Transfer matrix}


\begin{figure*}[htb]
\begin{center}
\includegraphics[width=0.99\linewidth]{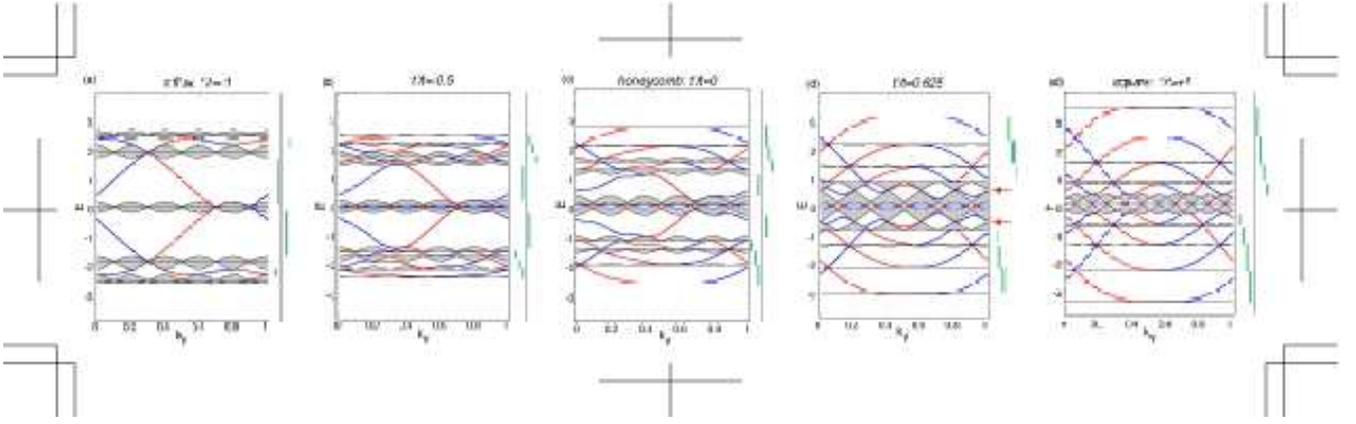}
\end{center}
\caption{
The energy spectra for the cylindrical system in Fig. \ref{f:zigzag_with_eq}
with the flux $\phi=1/5$ for 
(a) $t'/t=-1$ ($\pi$-flux),
(b) $t'/t=0.625$, 
(c) $t'/t=0$ (honeycomb), 
(d) $t'/t=0.5$, and 
(e) $t'/t=1$ (square).
Blue (red) lines are edge states localized at the bearded 
(zigzag) edges, while the bulk energy bands are shown as grey regions. 
The topological number, $I(E_{\rm F})$,  
when $E_{\rm F}$ is in a major gap 
in the negative energies are:
(a) $2$,  $-1$;  
(b) $1$, $2$, $3$,  $-1$; 
(c) $1$, $2$, $3$,  $-1$  
(d) $1$, $2$, $3$, (ill-defined)  
and
(e) $1$, $2$, $3$,  $4$.  
A topological transition accompanied with a discrete Chern number  
change by $\Delta c=-5$ (from $-4$ to +1)
at the 4-th major gap occurs between $t'/t=0.5$  $\to1$.
The green lines in the right are the topological numbers, $I(E_{\rm F})$,
obtained by counting the edge states.
}
\label{f:edges_1o5}
\end{figure*}

We follow the analysis 
in refs\cite{Hatsugai93a,Hatsugai93b,yh-qhe-review} 
to 
reduce the system to a one-dimensional model by making 
a partial Fourier transform of the fermion operators in $2$-direction,
\begin{alignat}{1}
c_{\circ,\bullet}(j) =&  
\frac {1}{\sqrt{L_2}} \sum_{k_2} e^{ik_2 j_2} c_{\circ,\bullet}(j_1,k_2). 
\end{alignat}
This yields a $k_2$-dependent series of one-dimensional Hamiltonian 
\begin{alignat}{1} 
&
H =  \sum_{k_2} 
H_{\rm 1D}(k_2),\nonumber \\
&
H_{\rm 1D}(k_2) =\sum_{j_1}\Big[
t_{\circ\bullet}(j_1,k_2)
c_\circ ^\dagger (j_1,k_2)
c_\bullet  (j_1,k_2) 
\nonumber\\
&  +
t_{\bullet\circ}(j_1,k_2)
c_\bullet ^\dagger  (j_1+1,k_2)
c_\circ   (j_1,k_2)\Big]
+ {\rm h.c.} 
\label{OneDimLocHam}
\end{alignat} 
where $k_2$-dependent hopping parameters are
\begin{alignat*}{1}
t_{\circ\bullet}(j_1,k_2)=& t\left(1+e^{ik_2-i 2\pi \phi j_1}\right) ,
\nonumber\\
t_{\bullet\circ}(j_1,k_2)=&t\left[1+ (t'/t)~e^{ik_2-i 2\pi \phi
 (j_1+1/2)}\right] .
\end{alignat*} 
Expanding one-particle eigenstates $|E,k_2 \rangle  $ with energy $E$
and momentum $k_2$ as
\begin{alignat*}{1} 
|E, k_2 \rangle =
 \sum_{j_1} \bigg[&
\psi_\bullet(E,j_1,k_2)c_{\bullet}^\dagger(j_1,k_2)|0\rangle
\nonumber\\ 
&+\psi_\circ(E,j_1,k_2)c_{\circ}^\dagger(j_1,k_2)|0\rangle \bigg],
\end{alignat*} 
we find that the Schr\"odinger equation 
$H_{\rm 1D}(k_2)|E,k_2\rangle=E|E,k_2\rangle$ is cast into a matrix form,
\begin{alignat*}{1} 
&\mvec
{\psi_\circ(j_1)}
{\psi_\bullet(j_1)}
= 
M_{\circ\bullet}(j_1)
\mvec
{\psi_\bullet(j_1)}
{\psi_\circ(j_1-1)},
\nonumber\\
&\mvec
{\psi_\bullet(j_1+1)}
{\psi_\circ(j_1)}
= 
M _{\bullet\circ}(j_1)
\mvec
{\psi_\circ(j_1)}
{\psi_\bullet(j_1)},
\end{alignat*} 
with
\begin{alignat}{1} 
&M_{\circ\bullet}(j_1) = 
\mat
{\frac E {t_{\circ\bullet} ^*(j_1)}}
{-\frac {t_{\bullet\circ}(j_1-1)} {t_{\circ\bullet} ^*(j_1)}}
{1}{0} ,
\nonumber\\
&M_{\bullet\circ}(j_1) =
\mat
{\frac E {t_{\bullet\circ} ^*(j_1)}}
{-\frac {t_{\circ\bullet}(j_1)} {t_{\bullet\circ} ^*(j_1)}}
{1}{0} .
\label{DefMcb}
\end{alignat} 
Therefore, we have 
\begin{alignat*}{1} 
&\mvec
{\psi_\bullet(j_1+1)}
{\psi_\circ(j_1)}
= 
M_{\rm t}(j_1)
\mvec
{\psi_\bullet(j_1)}
{\psi_\circ(j_1-1)}
\end{alignat*} 
with
\begin{alignat}{1}
M_{\rm t}(j_1)= 
M_{\bullet\circ}(j_1)M_{\circ\bullet}(j_1),
\label{TraMatt}
\end{alignat}
where every quantity is a function of $(E,j_1,k_2)$.  
They are a set of equations for $2 L_1-1$ variables 
$
 \psi_\bullet (j_x) (j_x=1,\cdots,L_x-1), 
 \psi_\circ (j_x) (j_x=1,\cdots,L_x) 
$ 
with open boundary conditions 
$\psi_\circ(0) = \psi_\circ(L_1) = 0$ 
(See Fig.\ref{f:zigzag_with_eq}).  
To impose this condition, we consider 
cylindrical systems with a zigzag edge at one end 
and a bearded (or Klein's) edge at the other 
as illustrated in Fig. \ref{f:zigzag_with_eq}.  
This is just a technical convention to apply the following 
transfer-matrix formalism.  
In terms of the energy spectrum, zigzag-bearded system 
has an $E=0$ edge state in zero magnetic field that 
extend over the entire Brillouin zone, as was 
first pointed out in \cite{kusakabe}.
In passing, we mention that
the local transformation 
\begin{alignat}{1} 
 \psi_\bullet(E,j_1) \to &  -\psi_\bullet(E,j_1) = \psi_\bullet(-E,j_1),
\nonumber\\
 \psi_\circ(E,j_1) \to &  \psi_\circ(E,j_1) = \psi_\circ(-E,j_1) ,
\end{alignat} 
yields the same Schrodinger equation  but with $-E$.
This is due to the bipartite symmetry of honeycomb: 
Eigenstates for $\pm E$ are paired with the same
amplitudes $|\psi_{\bullet,\circ}(m)|$
except for the zero energy.

Here note the periodicity 
$M_{s}(j_1+q)=M_{s}(j_1)$
for $s=\circ\bullet$, $s=\bullet\circ$, and hence
$M_{\rm t}(j_1+q)=M_{\rm t}(j_1)$. 
We can now define the transfer matrix for a unit period,
\begin{alignat}{1}
M(E,k_2)=\prod_{j_1=1}^qM_{\rm t}(E,j_1,k_2),
\label{TraMatUni}
\end{alignat}
then 
\begin{alignat}{1}
\left(
\begin{array}{c}
\psi_\bullet(q+1)\\
\psi_\circ(q)
\end{array}
\right)
=M
\left(
\begin{array}{c}
\psi_\bullet(1)\\
\psi_\circ(0)
\end{array}
\right).
\label{OnePer}
\end{alignat}
For $L_1=q \ell$ with an integer $\ell$ we have
\begin{alignat*}{1} 
\mvec
{\psi_\bullet(L_1+1)}
{\psi_\circ(L_1)}
=& 
M^\ell
\mvec
{\psi_\bullet(1)}
{\psi_\circ(0)},\quad
\end{alignat*}
with the boundary condition 
$\psi_\circ(0)=\psi_\circ(L_1)=0$ 
and $\psi_\bullet(1)=1 $, which 
is equivalent to an algebraic equation
$(M^\ell)_{21}(E)=0$ for $E$, which is a 
polynomial of order $2L_1-1$ and has as many real roots.

We can extract the edge states from the whole 
spectrum\cite{Hatsugai93a,Hatsugai93b}.  
Namely, $2q-1$ roots of $M_{21}(E) =0$, a 
polynomial equation of order $2q-1$, give the 
edge state energies, $E_j(k_2)$ for $j=1,\cdots,2q-1$.
By contrast, usual cylindrical systems where both edges 
are zigzag, bearded, or armchair do not allow such extraction.  
For example, in the case for 
zigzag edges, the boundary condition is replaced by
$\psi_\circ(0)=\psi_\bullet(L_1+1)=0$ in the present formulation.
Hence, the total spectra is determined by $(M^\ell)_{11}(E)=0$, giving 
$2L_1$ real roots. Edge states are contained in these spectra, but 
there is no simple way to directly extract them only. 

A remarkable feature here is that, on top of the edge states 
across adjacent Landau bands, 
there always exist exactly constant $E=0$ edge states,
since the number of the edge states are odd $2q-1$.  
Namely, a system with zigzag edges has 
a zero-energy edge mode in some region of $k_2$, 
a system with bearded edges in the other region, 
and the present system has the mode over the entire region.  
This is a simple extension of the discussion in the 
absence of a magnetic field. \cite{Ryu02}

We can further identify the spatial position of edge states: 
With 
$M_{11}(E_j)$,  we can show that 
\begin{alignat*}{1} 
|M_{11}(E_j)| < 1 :& \text{ localized at the zigzag edge}, 
\\
|M_{11}(E_j)| > 1 :& \text{ localized at the bearded edge} .
\end{alignat*} 
With Laughlin's argument\cite{Laughlin81}, 
supplemented by the above 
behavior of the edge state (as a function of the momentum $k_2$), 
one can identify a topological number $I(E_{\rm F})$, 
the number of electrons carried by 
Laughlin's adiabatic procedure, for each edge state branch when
the chemical potential $E_{\rm F}$ 
traverses the branch. \cite{Laughlin81,Hatsugai93a,Hatsugai93b,yh-qhe-review}


Having determined the spectra 
for the edge states with the transfer matrix, 
we can now relate the edge states to the Bloch functions
of the bulk systems: 
Let us relax the open boundary condition, and 
impose instead a periodic boundary condition 
$H(j_1+q,k_2)=H(j_1,k_2)$ on the local Hamiltonian (\ref{OneDimLocHam}).  
Then the Bloch theorem for the bulk state leads to 
$\psi_{\bullet}(j_1+q) = e^{i  k_1 q}\psi_{\bullet}(j_1)$
and 
$\psi_{\circ}(j_1+q) = e^{i  k_1 q}\psi_{\circ}(j_1)$, or
\begin{alignat*}{1} 
\mvec
{\psi_\bullet(j_1+q+1)}
{\psi_\circ(j_1+q)}
&= 
e^{ik_1q}
\mvec
{\psi_\bullet(j_1+1)}
{\psi_\circ(j_1)}.
\end{alignat*} 
Comparing this with Eq. (\ref{OnePer}), we see that 
the eigenvalue of the transfer matrix $M$ is $e^{ik_1q}$, i.e, 
just a phase factor.
Therefore, the energy bands are specified by the following condition for
the eigenvalue $\rho$ of the transfer matrix $M$,
$
\det (\rho 1- M)= 0
$, with $ |\rho| =1$.
As shown in the App. \ref{app1}, the condition for the energy band is given by
$
|{\rm Tr \,} M  | \le 2$.
\begin{figure}[htb]
\begin{center}
\includegraphics[width=0.7\linewidth]{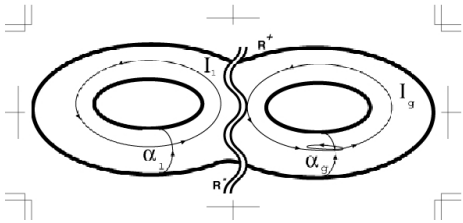}
\end{center}
\caption{
A Riemann surface  $\Sigma_{2q-1}(k_2)$ for 
the Bloch function which is a complex energy surface for 
defining $\rho(E)$ consistently.
Its genus is $2q-1$, which coincides with the number of energy gaps. 
}
\label{f:RS}
\end{figure}

The eigenvalue of the transfer matrix is also written as
$
\rho = 
\frac {1}{2} \left\{
 {\rm Tr \,} M -
\left[({\rm Tr \,} M)^2- 4 \det M \right]^{1/2} 
\right\}
$,
where we have to fix the branch of the square root. 
Then the Riemann surface,
$\Sigma_{2q-1}$,  of the Bloch function is given by 
that of the eigenvalue $\rho(E)$ and
its genus is generically $2q-1$ 
which are the number of energy gaps (Fig.\ref{f:RS}). 
With this Riemann surface 
and the arguments in Refs\cite{Hatsugai93a,Hatsugai93b,yh-qhe-review},
one can establish the bulk-edge correspondence as 
\begin{alignat*}{1} 
c_{\rm F}(E_{\rm F}) =& 
I(E_{\rm F}),
\end{alignat*} 
where $c_{\rm F}(E_{\rm F})$ is the Chern number obtained in 
section III, while $I(E_{\rm F})$ is the topological number for edge states 
which corresponds to the number of intersections on
the Riemann surface $\Sigma_{2q-1}$.

\subsection{Edge-bulk equivalence and topological transition}

Before we discuss the experimental situation, 
let us look at the case of strong magnetic fields with 
$\phi=1/q$ with $q=O(1)$. 
This case, while unrealistic, is useful for heuristic purposes.
In Fig. \ref{f:edges_1o5},
the spectra of the cylindrical systems are shown for a 
relatively large $\phi=1/5$.  
The shaded regions are Landau bands, while 
the blue (red) curves across the bands correspond to edge states 
localized at the bearded (zigzag) edges.  
It should be noted that we always find a dispersionless edge state at
the zero energy, which is due to bipartite symmetry. 
This is the state employed for the boundary magnetic moment, 
but does not contribute to the topological number, since
the number is ill-defined due to the gap-closing at the zero energy.

For the honeycomb lattice there are 9 bands for $\phi=1/5$, as observed in
Fig. \ref{f:edges_1o5} (c). 
By counting the number of the edge states, we can see that 
the topological numbers $I(E_{\rm F})$ 
are 1, 2, 3, $-1$, +1, $-3$, $-2$ and $-1$, when 
$E_{\rm F}$ lies in the $j$th major gap 
for $j=1, \cdots, 8$, respectively.
Note that gap-closing occurs at the zero energy, and 
the increment of the topological number across these degenerate bands is 
$\Delta I=2$ which corresponds to the Dirac-Landau level quantization of
the Hall conductivity discussed in Sec. \ref{s:Bul}.

\begin{figure*}[htb]
\begin{center}
\includegraphics[width=0.99\linewidth]{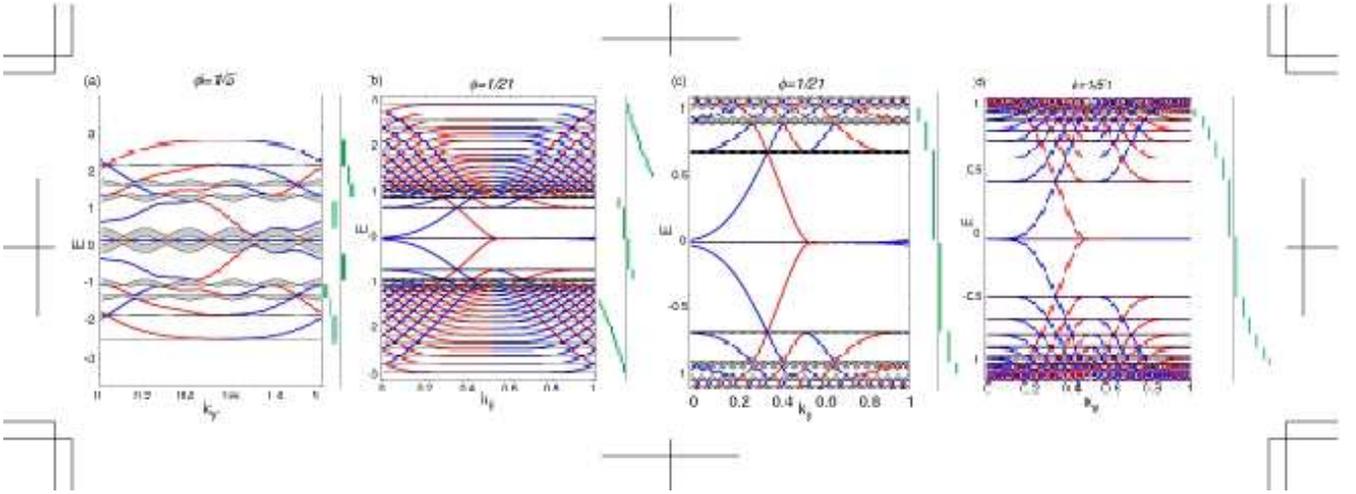}
\end{center}
\caption{
The energy spectra of the honeycomb lattice 
with zigzag and bearded edges for 
(a) $\phi=1/5$
and (b) $\phi=1/21$, 
(c) is a blowup of the (b), and
(d) is for $\phi=1/51$.
The topological number $I(E_{\rm F})$  
when $E_{\rm F}$ is in a major energy gaps is also indicated 
next to the vertical axis.
}
\label{f:edges_honey_full}
\end{figure*}


If we move on to $\pi$-flux lattice ($t'/t:0\to -1$), 
the Dirac fermion level with $\Delta I=2$  at the zero energy 
persists (Fig. \ref{f:edges_1o5} (a) and (b)). This is 
the {\it edge-bulk} correspondence version of
 the topological equivalence (I) in
Sec. \ref{s:TopEqu}.  
If we move on to square lattice ($t'/t:0\to1$), on the other hand, 
we can observe a topological change in these bands:
At $t'/t=1$, the topological numbers $I(E_{\rm F})$ reads 
1, 2, 3, 4, $-4$, $-3$, $-2$, $-1$ versus $E_{\rm F}$, 
which dictates that $E=0$ is now a van-Hove energy with 
$|\Delta I|=8\gg 1$.  
This corresponds to the 
fact that van-Hove energies merge to wash out the Dirac-Landau 
region precisely at $t'/t=1$, so the square lattice is singular 
in this respect.  
The merging of 
the relevant Landau bands (4th-6th for $\phi=1/5$) is clearly seen in 
the spectra at $t'/t=0.625$ shown in Fig. \ref{f:edges_1o5}(d), where 
the topological number $-5$
moves to the zero energy level from the levels just below and above it.

We next investigate
the second equivalence 2 given in section III  
via similar observations.
Let us start with the honeycomb lattice.  The increment in the
topological number is $\Delta I=1$ in the band edge regions, 
the conventional Fermi-Landau quantization. 
This feature indeed survives in the adiabatic process
$t'/t:0\to1$ in this region. This is the topological equivalence 2. 
On the other hand, for $t'/t:0\to-1$ these bands merge 
{\it in pairs} to produce doubled Chern numbers, 
which corresponds to the fermion doubling due to the multiple 
band extrema discussed in \ref{s:CheNumDvsF}.

In passing, we mention that 
one can confirm the particle-hole symmetry 
i.e, the bipartite symmetry in the topological numbers,
\begin{alignat}{1} 
I(E_{\rm F}) =& -I(-E_{\rm F}).
\end{alignat} 

\subsection{From strong to weak magnetic fields}

Having identified the strong-field behavior, 
let us next 
show how the spectra changes when the magnetic field 
has smaller, more realistic values. 
In Fig. \ref{f:edges_honey_full}, we show the spectra 
for the honeycomb lattice.  
The weaker the external magnetic field becomes, 
the narrower the gaps as well as the band widths become: 
For $\phi=1/31$, the bands look like Landau levels, 
but the edge states continue to exhibit characteristic behavior
around the zero energy region 
(1, 3, 5, ... edge branches for the consecutive gaps), 
and this region is separated by
the van-Hove energies $\sim \pm1$, outside of which 
we recover the free fermionic topological numbers with $\Delta I=1$.
As discussed above,
in strong magnetic fields such as $\phi=1/5$, 
only two bands across the 
zero-energy are Dirac-Landau levels, while 
the next bands with energy $\sim\pm 1$ are van-Hove bands.
Decreasing magnetic field makes 
the number of Dirac fermion bands increases: 
The number of edge states in Fig. \ref{f:edges_honey_full} 
is precisely in accord with this.

\section{Summary and discussions} \label{s:Sum}
In summary, we have shown that the Landau levels are divided by the van Hove
singularities into two regimes:
One is effectively described by Dirac particles, and the other by
ordinary finite-mass fermions. 
Remarkably, this persists as we convert the lattice into 
square or $\pi$-flux lattices.   
We have confirmed this both from the 
Chern number argument on Hofstadter's diagram and 
by the bulk-edge correspondence. 

The van Hove singularities have turn out to play a key role 
in separating the Dirac-Landau and Fermi-Landau regions.  
Thus the anomalous QHE is not unique to the honeycomb lattice, 
but continuously shared by a class of lattices (except by 
the square lattice).  The Dirac-Landau behavior can also occur 
when there are multiple extrema in the dispersion, 
as in the region between honeycomb and $\pi$-flux lattices.

The properties revealed here for the whole spectrum of the honeycomb lattice 
may be experimentally observable if the chemical potential can be 
varied over a wide range.  
In real graphene samples there may be disorder, in which case 
we are talking about a dirty Hofstadter system.  
However, non-monotonic behaviors should survive the disorder as far 
as the degree of disorder is not too large, as has been 
indicated by a numerical calculation for a dirty Hofstadter system\cite{aoki92}.
As for the $\pi$-flux lattice, 
the model effectively describes the excitation spectrum of 
d-wave superconductors, so there may exist a relevance.

\begin{acknowledgements}
This work was in part supported by 
Grants in Aids from the Japanese Ministry of Education 
(for Creative Scientific Research Project 
"Quantum properties of organic-inorganic complex structures" for HA, 
 Grant-in-Aid for Scientific Research  
(No.17540347) from JSPS and on Priority Areas (Grant No.18043007) from MEXT for YH
and 
 Grant-in-Aid for Scientific Research  
(No.18540365) from JSPS for TF.
YH was also supported in part by the Sumitomo foundation. 
\end{acknowledgements}

\appendix

\section{Momentum representation of the Hamiltonian} \label{s:AppMom}
We derive the Hamiltonian for the honeycomb lattice 
in the momentum representation for the flux per hexagon $\phi=p/q$, 
where $p$ and $q$ are mutually prime integers.  
We label the unit cell as  
$\mib{j}=j_1\mib{e}_1+j_2\mib{e}_2$.
Let $\mib{r}_\mu$ be
the reciprocal vectors (divided by $2\pi$) satisfying 
$\mib{e}_\mu\cdot \mib{r}_\nu=\delta_{\mu\nu}$.  
In an external magnetic field, it is convenient to introduce a 
larger unit cell along coordinate 1 as
$J=J_1(qe_1)+J_2 e_2$. 
In this unit cell, the fermi operators are denoted as
$c_{sm}(J)$ with $s=\bullet,\circ$ and $\quad m=1,\cdots,q$.
Let $\mib{K}$ be the corresponding momentum defined by
$\mib{K}=K_1(\mib{r}_1/q)+K_2\mib{r}_2$.
Then the Fourier transformation can be defined as
\begin{alignat*}{1}
c_{sm}(J)=\int_0^{2\pi}\frac{d {K}}{(2\pi)^2}
e^{i {K}\cdot {J}}c_{sm}({K}) .
\end{alignat*}
Note that $K_\mu\in[0,2\pi]$. 
Introduce $K_1/q=k_1$ and $K_2=k_2$, where $k_\mu$ denotes the 
momentum in zero magnetic field.
It is convenient to introduce operators 
$\bm c^\dagger({k})=\left(c_{\bullet1}^\dagger({k}), c_{\circ1}^\dagger({k}), 
c_{\bullet2}^\dagger({k}), c_{\circ2}^\dagger({k}),\hdots ,
c_{\bullet q}^\dagger({k}), c_{\circ q}^\dagger({k})
\right)$.
With these we obtain the Hamiltonian in the momentum representation
Eq. (\ref{HamMom}) as
\begin{alignat}{1}
&
h (k)= 
\left(
\begin{array}{ccccc}
 d_1
&  
 f_1
& 
& 
&  
e^{-iq k_1} f_q ^\dagger
\\
 f_1 ^\dagger
& 
 d_2
&   
 f_2
& 
&  
\\
& 
& 
 \ddots  
& 
&  
\\
&   
& 
& 
 d_{q-1}
&   
 f_{q-1}
\\
 e^{+iq k_1}  f_q
&
&    
&  
 f_{q-1} ^\dagger
& 
 d_{q}
\end{array}
\right),
\label{HamEle}
\end{alignat}
where
\begin{alignat*}{1}
&
d_j(k_2)
=  
t\mat
{0}{ 1+e^{-ik_2+i2\pi \phi j}}
{ 1+e^{ik_2-i2\pi \phi j}}{0},
\nonumber\\
&
f_j (k_2)
=  
t\mat
{0}{0}
{
1+ (t'/t)e^{-ik_2+i2\pi\phi( j+1/2)}
}{0} .
\end{alignat*} 

\section{Gauge Transformation and the Transfer Matrix}
\label{app1}

In the transfer matrix formalism, the energy bands of the bulk system
can be determined by the condition
\begin{alignat}{1}
\rho^2-\rho{\rm Tr}M+\det M=0
\end{alignat}
that satisfies $|\rho|=1$.

Since the transfer matrix has complex matrix elements, 
let us separate the hopping parameters into the magnitude and the phase as
\begin{alignat*}{1}
&
t_{s}(j_1,k_2)=r_{s}(j_1,k_2)e^{i\phi_{s}(j_1,k_2)} ,
\quad
s=\circ\bullet, \bullet\circ ,
\end{alignat*}
where $r_s$ is a real non-negative parameter. 
Obviously, $r_s(j_1+q,k_2)=r_s(j_1,k_2)$.
In what follows, 
the $k_2$ dependences are suppressed for simplicity.
\begin{widetext}
Then the matrices in Eq. (\ref{DefMcb}) can be expressed as
\begin{alignat*}{1}
&
M_{\circ\bullet}(j_1)=
\left(
\begin{array}{cc}
e^{i\phi_{\circ\bullet}(j_1)+i\phi_{\bullet\circ}(j_1-1)}
&
0\\
0
&
e^{i\phi_{\bullet\circ}(j_1-1)}
\end{array}
\right)
\left(
\begin{array}{cc}
\frac{E}{r_{\circ\bullet}(j_1)}
&
-\frac{r_{\bullet\circ}(j_1-1)}{r_{\circ\bullet}(j_1)}
\\
1
&
0
\end{array}
\right) 
\left(
\begin{array}{cc}
e^{-i\phi_{\bullet\circ}(j_1-1)}
&
0\\
0
&
1
\end{array}
\right),
\nonumber\\
&
M_{\bullet\circ}(j_1)=
\left(
\begin{array}{cc}
e^{i\phi_{\circ\bullet}(j_1)+i\phi_{\bullet\circ}(j_1)}
&
0\\
0
&
e^{i\phi_{\circ\bullet}(j_1)}
\end{array}
\right)
\left(
\begin{array}{cc}
\frac{E}{r_{\bullet\circ}(j_1)}
&
-\frac{r_{\circ\bullet}(j_1)}{r_{\bullet\circ}(j_1)}
\\
1
&
0
\end{array}
\right) 
\left(
\begin{array}{cc}
e^{-i\phi_{\circ\bullet}(j_1)}
&
0\\
0
&
1
\end{array}
\right).
\end{alignat*}
By multiplying the above two matrices, 
we can see that the transfer matrix (\ref{TraMatt}) can be written as
\begin{alignat*}{1}
M_{\rm t}(j_1)&=
\left(
\begin{array}{cc}
e^{i\phi_{\circ\bullet}(j_1)+i\phi_{\bullet\circ}(j_1)}
&
0\\
0
&
e^{i\phi_{\circ\bullet}(j_1)}
\end{array}
\right)
\left(
\begin{array}{cc}
\frac{E}{r_{\bullet\circ}(j_1)}
&
-\frac{r_{\circ\bullet}(j_1)}{r_{\bullet\circ}(j_1)}
\\
1
&
0
\end{array}
\right) 
\left(
\begin{array}{cc}
\frac{E}{r_{\circ\bullet}(j_1)}
&
-\frac{r_{\bullet\circ}(j_1-1)}{r_{\circ\bullet}(j_1)}
\\
1
&
0
\end{array}
\right) 
\left(
\begin{array}{cc}
1
&
0\\
0
&
e^{i\phi_{\bullet\circ}(j_1-1)}
\end{array}
\right) 
\nonumber\\
&=U(j_1+1)\widetilde M_{\rm t}(j_1)U^{-1}(j_{1}),
\end{alignat*}
where $\widetilde M_{\rm t}$ is a real matrix defined by 
the product of the second and third
matrices in the r.h.s. of the first line, and $U(j_1)$ 
is a diagonal matrix defined by
\begin{alignat*}{1}
U(j_1)&=
\exp i \left(
\begin{array}{cc}
\displaystyle{\sum_{j_1'}^{j_1-1}}
\left[
\phi_{\circ\bullet}(j_1')+\phi_{\bullet\circ}(j_1')
\right]
&
0\\
0
&
\displaystyle{\sum_{j_1'}^{j_1-1}}
\phi_{\circ\bullet}(j_1')
+\displaystyle{\sum_{j_1'}^{j_1-2}}
\phi_{\bullet\circ}(j_1')
\end{array}
\right) .
\end{alignat*}
\end{widetext}
Here,
the sum over $j_1'$ starts from some integer, say, $-1$. The matrix
$U(j_1)$ is nothing but a local gauge transformation for the wave functions 
$\phi_{\circ}(j_1)$ and $\psi_\bullet(j_1)$. 
Therefore, we arrive at an expression for the unit cell transfer
matrix (\ref{TraMatUni}),
\begin{alignat*}{1}
M=U(q+1)\widetilde M U^{-1}(1) ,
\end{alignat*}
where $\widetilde M$ is a real matrix defined by 
$\widetilde M\equiv \prod_{j_1=1}^q \widetilde M_{\rm t}(j_1)$.
Note that
\begin{alignat*}{1}
U^{-1}(1)U(q+1) =&   e^{i\Theta}{\bf 1} ,
\end{alignat*}
where $\Theta$ is defined by 
$\Theta \equiv  \sum_{j_1=1}^q\sum_s \phi_{s}(j_1)=
\sum_{j_1=1}^q {\rm Arg \,}  {t_{\bullet\circ}(j_1)}
{t_{\circ\bullet}(j_1)}$ (mod $2\pi$).
Finally, we have
\begin{alignat*}{1}
&
\det M=e^{2i\Theta} ,
\nonumber\\
&
{\rm Tr} M=e^{i\Theta}{\rm Tr} \widetilde M ,
\end{alignat*}
where we have used $\det \widetilde M=1$ in the first line.
It should be stressed again that ${\rm Tr}\widetilde M$ is real.
Thus we conclude that the quadratic equation 
has 
solutions that satisfy $|\rho|=1$ if 
$|{\rm Tr}\widetilde M|=|{\rm Tr}M|\le2$.

As a function of the energy, 
the eigenvalue of the transfer matrix is written as
\begin{alignat*}{1} 
\rho =& 
\frac {1}{2} \left\{
 {\rm Tr \,} M -
\left[({\rm Tr \,} M)^2- 4 \det M \right]^{1/2} 
\right\},
\end{alignat*} 
where we have to fix the branch of the square root. 
Then the Riemann surface,
$\Sigma_{2q-1}$,  of the Bloch function is given by 
that of the eigenvalue $\rho(E)$ and
its genus is generically $2q-1$ 
which are the number of energy gaps.


\begin{thebibliography}{99}
\expandafter\ifx\csname natexlab\endcsname\relax\def\natexlab#1{#1}\fi
\expandafter\ifx\csname bibnamefont\endcsname\relax
  \def\bibnamefont#1{#1}\fi
\expandafter\ifx\csname bibfnamefont\endcsname\relax
  \def\bibfnamefont#1{#1}\fi
\expandafter\ifx\csname citenamefont\endcsname\relax
  \def\citenamefont#1{#1}\fi
\expandafter\ifx\csname url\endcsname\relax
  \def\url#1{\texttt{#1}}\fi
\expandafter\ifx\csname urlprefix\endcsname\relax\def\urlprefix{URL }\fi
\providecommand{\bibinfo}[2]{#2}
\providecommand{\eprint}[2][]{\url{#2}}

\bibitem[{\citenamefont{Eguchi et~al.}(1980)\citenamefont{Eguchi, P.B. Gilkey,
  and A.J. Hanson}}]{eguchi80}
\bibinfo{author}{\bibfnamefont{T. }~\bibnamefont{Eguchi}},
  \bibinfo{author}{\bibnamefont{P.B. Gilkey}}, \bibnamefont{and}
  \bibinfo{author}{\bibnamefont{A.J. Hanson}}, \bibinfo{journal}{Phys.\ Rep.}
  \textbf{\bibinfo{volume}{66}}, \bibinfo{pages}{213} (\bibinfo{year}{1980}).

\bibitem[{\citenamefont{S.B.Treiman et~al.}(1985)\citenamefont{S.B. Treiman, 
R. Jackiw, B. Zumino, and E. Witten}}]{current85}
\bibinfo{editor}{\bibnamefont{S.B. Treiman}},
  \bibinfo{editor}{\bibfnamefont{R. }~\bibnamefont{Jackiw}},
  \bibinfo{editor}{\bibnamefont{B. Zumino}}, \bibnamefont{and}
  \bibinfo{editor}{\bibfnamefont{E. }~\bibnamefont{Witten}}, eds.,
  \emph{\bibinfo{title}{Current Algebra and Anomaly}}
  (\bibinfo{publisher}{World Scientific}, \bibinfo{year}{1985}).

\bibitem[{\citenamefont{Semenoff}(1984)}]{Seme84}
\bibinfo{author}{\bibfnamefont{G. }~\bibnamefont{Semenoff}},
  \bibinfo{journal}{Phys. \ Rev.\ Lett.} \textbf{\bibinfo{volume}{53}},
  \bibinfo{pages}{2449} (\bibinfo{year}{1984}).

\bibitem[{\citenamefont{Haldane}(1987)}]{Hal87}
\bibinfo{author}{\bibfnamefont{F.~D.~M.} \bibnamefont{Haldane}},
  \bibinfo{journal}{Phys. \ Rev.\ Lett.} \textbf{\bibinfo{volume}{61}},
  \bibinfo{pages}{2015} (\bibinfo{year}{1987}).

\bibitem[{\citenamefont{Lee}(1993)}]{Lee93}
\bibinfo{author}{\bibfnamefont{P.~A.} \bibnamefont{Lee}},
  \bibinfo{journal}{Phys. \ Rev.\ Lett.} \textbf{\bibinfo{volume}{71}},
  \bibinfo{pages}{1887} (\bibinfo{year}{1993}).

\bibitem{NTW94}A. A. Nersesyan, A. M. Tsvelik, and F. Wenger, Phys. Rev. Lett. 72, 2628 (1994).

\bibitem[{\citenamefont{Hasegawa et~al.}(1989)\citenamefont{Hasegawa, Lederer,
  Rice, and Wiegmann}}]{Hase89}
\bibinfo{author}{\bibfnamefont{Y.}~\bibnamefont{Hasegawa}},
  \bibinfo{author}{\bibfnamefont{P.}~\bibnamefont{Lederer}},
  \bibinfo{author}{\bibfnamefont{T.~M.} \bibnamefont{Rice}}, \bibnamefont{and}
  \bibinfo{author}{\bibfnamefont{P.~B.} \bibnamefont{Wiegmann}},
  \bibinfo{journal}{Phys. \ Rev.\ Lett.} \textbf{\bibinfo{volume}{63}},
  \bibinfo{pages}{907} (\bibinfo{year}{1989}).

\bibitem[{\citenamefont{Wen et~al.}(1989)\citenamefont{Wen, Wilczek, and
  Zee}}]{Wen89wwz}
\bibinfo{author}{\bibfnamefont{X.~G.} \bibnamefont{Wen}},
  \bibinfo{author}{\bibfnamefont{F.}~\bibnamefont{Wilczek}}, \bibnamefont{and}
  \bibinfo{author}{\bibfnamefont{A.}~\bibnamefont{Zee}},
  \bibinfo{journal}{Phys. \ Rev.\ B} \textbf{\bibinfo{volume}{39}},
  \bibinfo{pages}{11413} (\bibinfo{year}{1989}).

\bibitem[{\citenamefont{Fisher and Fradkin}(1984)}]{Fish84}
\bibinfo{author}{\bibfnamefont{M.~P.~A.} \bibnamefont{Fisher}}
  \bibnamefont{and} \bibinfo{author}{\bibfnamefont{E.}~\bibnamefont{Fradkin}},
  \bibinfo{journal}{Nucl.\ Phys.\ B} \textbf{\bibinfo{volume}{251}},
  \bibinfo{pages}{457} (\bibinfo{year}{1984}).


\bibitem[{\citenamefont{Ohgushi et~al.}(2000)\citenamefont{Ohgushi, Murakami,
  and Nagaosa}}]{Ohgushi00}
\bibinfo{author}{\bibfnamefont{K.}~\bibnamefont{Ohgushi}},
  \bibinfo{author}{\bibfnamefont{S.}~\bibnamefont{Murakami}}, \bibnamefont{and}
  \bibinfo{author}{\bibfnamefont{N.}~\bibnamefont{Nagaosa}},
  \bibinfo{journal}{Phys.\ Rev.\ B} \textbf{\bibinfo{volume}{62}},
  \bibinfo{pages}{R6065} (\bibinfo{year}{2000}).

\bibitem[{\citenamefont{Kane and Mele}(2005)}]{KM05}
\bibinfo{author}{\bibfnamefont{C.~L.} \bibnamefont{Kane}} \bibnamefont{and}
  \bibinfo{author}{\bibfnamefont{E.~J.} \bibnamefont{Mele}},
  \bibinfo{journal}{Phys.\ Rev.\ Lett.} \textbf{\bibinfo{volume}{95}},
  \bibinfo{pages}{146802} (\bibinfo{year}{2005}).

\bibitem[{\citenamefont{Ludwig et~al.}(1993)\citenamefont{Ludwig, Fisher,
  Shankar, and Grinstein}}]{Lud93}
\bibinfo{author}{\bibfnamefont{A.~W.~W.} \bibnamefont{Ludwig}},
  \bibinfo{author}{\bibfnamefont{M.~P.~A.} \bibnamefont{Fisher}},
  \bibinfo{author}{\bibfnamefont{R.}~\bibnamefont{Shankar}}, \bibnamefont{and}
  \bibinfo{author}{\bibfnamefont{G.}~\bibnamefont{Grinstein}},
  \bibinfo{journal}{Phys. \ Rev.\ B} \textbf{\bibinfo{volume}{50}},
  \bibinfo{pages}{7526} (\bibinfo{year}{1993}).

\bibitem[{\citenamefont{Hatsugai and Kohmoto}(1990)}]{Hat90}
\bibinfo{author}{\bibfnamefont{Y.}~\bibnamefont{Hatsugai}} \bibnamefont{and}
  \bibinfo{author}{\bibfnamefont{M.}~\bibnamefont{Kohmoto}},
  \bibinfo{journal}{Phys. \ Rev.\ B} \textbf{\bibinfo{volume}{42}},
  \bibinfo{pages}{8282} (\bibinfo{year}{1990}).

\bibitem[{\citenamefont{Zheng and Ando}(2002)}]{ZhengAndo}
\bibinfo{author}{\bibfnamefont{Y.}~\bibnamefont{Zheng}} \bibnamefont{and}
  \bibinfo{author}{\bibfnamefont{T.}~\bibnamefont{Ando}},
  \bibinfo{journal}{Phys. \ Rev.\ B} \textbf{\bibinfo{volume}{65}},
  \bibinfo{pages}{245420} (\bibinfo{year}{2002}).

\bibitem[{\citenamefont{Gusynin and Sharapov}(2005)}]{Gus05}
\bibinfo{author}{\bibfnamefont{V.~P.} \bibnamefont{Gusynin}} \bibnamefont{and}
  \bibinfo{author}{\bibfnamefont{S.}~\bibnamefont{Sharapov}},
  \bibinfo{journal}{Phys. \ Rev.\ Lett.} \textbf{\bibinfo{volume}{95}},
  \bibinfo{pages}{146801} (\bibinfo{year}{2005}).

\bibitem{Nov05}
K. S.~Novoselov, Nature, {\bf 438}, 197 (2005); Nature Physics, {\bf 2}, 177 (2006).


\bibitem{Zha05}
\bibinfo{author}{\bibfnamefont{Y.}~\bibnamefont{Zhang}},
  \bibinfo{author}{\bibfnamefont{Y.-W.} \bibnamefont{Tan}},
  \bibinfo{author}{\bibfnamefont{H.~L.} \bibnamefont{Stormer}},
  \bibnamefont{and} \bibinfo{author}{\bibfnamefont{P.}~\bibnamefont{Kim}},
  \bibinfo{journal}{Nature}, {\bf 438}, 201  (\bibinfo{year}{2005}).

\bibitem[{\citenamefont{Ando}(2005)}]{Ando05}
\bibinfo{author}{\bibfnamefont{T.}~\bibnamefont{Ando}}, \bibinfo{journal}{J.\
  Phys. \ Soc.\ Jpn.} \textbf{\bibinfo{volume}{74}}, \bibinfo{pages}{777}
  (\bibinfo{year}{2005}).

\bibitem[{\citenamefont{Fujita et~al.}(1996)\citenamefont{Fujita, Wakabayashi,
  Nakada, and Kusakabe}}]{Fuji96}
\bibinfo{author}{\bibfnamefont{M.}~\bibnamefont{Fujita}},
  \bibinfo{author}{\bibfnamefont{K.}~\bibnamefont{Wakabayashi}},
  \bibinfo{author}{\bibfnamefont{K.}~\bibnamefont{Nakada}}, \bibnamefont{and}
  \bibinfo{author}{\bibfnamefont{K.}~\bibnamefont{Kusakabe}},
  \bibinfo{journal}{J.\ Phys.\ Soc.\ Jpn.} \textbf{\bibinfo{volume}{65}},
  \bibinfo{pages}{1920} (\bibinfo{year}{1996}).

\bibitem[{\citenamefont{Wakabayashi et~al.}(1998)\citenamefont{Wakabayashi,
  adn, Ajiki, and Sigrist}}]{Waka98}
\bibinfo{author}{\bibfnamefont{K.}~\bibnamefont{Wakabayashi}},
  \bibinfo{author}{\bibfnamefont{M.} \bibnamefont{Fujita}},
  \bibinfo{author}{\bibfnamefont{H.}~\bibnamefont{Ajiki}}, \bibnamefont{and}
  \bibinfo{author}{\bibfnamefont{M.}~\bibnamefont{Sigrist}},
  \bibinfo{journal}{Phys.\ Rev.\ B} \textbf{\bibinfo{volume}{59}},
  \bibinfo{pages}{8271} (\bibinfo{year}{1999}).

\bibitem[{\citenamefont{Berry}(1984)}]{Berry84}
\bibinfo{author}{\bibfnamefont{M. V. }~\bibnamefont{Berry}}, 
  \bibinfo{journal}{Proc.\ Roy. \ Soc.\ London.\ A} 
  \textbf{\bibinfo{volume}{392}}, \bibinfo{pages}{45}
  (\bibinfo{year}{1984}).

\bibitem[{\citenamefont{Simon}(1983)}]{Simon83}
\bibinfo{author}{\bibfnamefont{B. }~\bibnamefont{Simon}}, 
  \bibinfo{journal}{Phys.\ Rev. \ Lett.} 
  \textbf{\bibinfo{volume}{51}}, \bibinfo{pages}{2167}
  (\bibinfo{year}{1983}).

\bibitem[{\citenamefont{Wen}(1989)}]{Wen89}
\bibinfo{author}{\bibfnamefont{X.~G.} \bibnamefont{Wen}},
  \bibinfo{journal}{Phys. \ Rev.\ B} \textbf{\bibinfo{volume}{40}},
  \bibinfo{pages}{7387} (\bibinfo{year}{1989}).

\bibitem[{\citenamefont{Shapere and Wilczek}(1989)}]{Shapere89}
\bibinfo{editor}{\bibfnamefont{A.}~\bibnamefont{Shapere}} \bibnamefont{and}
  \bibinfo{editor}{\bibfnamefont{F.}~\bibnamefont{Wilczek}}, eds.,
  \emph{\bibinfo{title}{Geometric Phases in Physics}}
  (\bibinfo{publisher}{World Scientific}, \bibinfo{year}{1989}).

\bibitem[{\citenamefont{Hatsugai}(2004{\natexlab{a}})}]{Hat04}
\bibinfo{author}{\bibfnamefont{Y.}~\bibnamefont{Hatsugai}},
  \bibinfo{journal}{J.\ Phys. \ Soc.\ Jpn.} \textbf{\bibinfo{volume}{73}},
  \bibinfo{pages}{2604} (\bibinfo{year}{2004}{\natexlab{a}}).

\bibitem[{\citenamefont{Hatsugai}(2004{\natexlab{b}})}]{Hat05}
\bibinfo{author}{\bibfnamefont{Y.}~\bibnamefont{Hatsugai}},
  \bibinfo{journal}{J.\ Phys. \ Soc.\ Jpn.} \textbf{\bibinfo{volume}{74}},
  \bibinfo{pages}{1374} (\bibinfo{year}{2004}{\natexlab{b}}).

\bibitem[{\citenamefont{Laughlin}(1981)}]{Laughlin81}
\bibinfo{author}{\bibfnamefont{R.~B.} \bibnamefont{Laughlin}},
  \bibinfo{journal}{Phys.\ Rev.\ B} \textbf{\bibinfo{volume}{23}},
  \bibinfo{pages}{5632} (\bibinfo{year}{1981}).

\bibitem[{\citenamefont{Halperin}(1982)}]{Halperin82}
\bibinfo{author}{\bibfnamefont{B.~I.} \bibnamefont{Halperin}},
  \bibinfo{journal}{Phys.\ Rev.\ B} \textbf{\bibinfo{volume}{25}},
  \bibinfo{pages}{2185} (\bibinfo{year}{1982}).

\bibitem[{\citenamefont{Hatsugai}(1993{\natexlab{a}})}]{Hatsugai93a}
\bibinfo{author}{\bibfnamefont{Y.}~\bibnamefont{Hatsugai}},
  \bibinfo{journal}{Phys.\ Rev.\ B} \textbf{\bibinfo{volume}{48}},
  \bibinfo{pages}{11851} (\bibinfo{year}{1993}{\natexlab{a}}).

\bibitem[{\citenamefont{Hatsugai}(1993{\natexlab{b}})}]{Hatsugai93b}
\bibinfo{author}{\bibfnamefont{Y.}~\bibnamefont{Hatsugai}},
  \bibinfo{journal}{Phys.\ Rev. \ Lett.} \textbf{\bibinfo{volume}{71}},
  \bibinfo{pages}{3697} (\bibinfo{year}{1993}{\natexlab{b}}).

\bibitem[{\citenamefont{Witten}(1982)}]{Wit82}
\bibinfo{author}{\bibfnamefont{E.}~\bibnamefont{Witten}},
  \bibinfo{journal}{Phys. \ Lett.} \textbf{\bibinfo{volume}{117B}},
  \bibinfo{pages}{324} (\bibinfo{year}{1982}).

\bibitem[{\citenamefont{Niemi and Semenoff}(1983)}]{Niem83}
\bibinfo{author}{\bibfnamefont{A.~J.} \bibnamefont{Niemi}} \bibnamefont{and}
  \bibinfo{author}{\bibfnamefont{G.}~\bibnamefont{Semenoff}},
  \bibinfo{journal}{Phys. \ Rev.\ Lett.} \textbf{\bibinfo{volume}{51}},
  \bibinfo{pages}{2077} (\bibinfo{year}{1983}).


\bibitem[{\citenamefont{Ryu and Hatsugai}(2002)}]{Ryu02}
\bibinfo{author}{\bibfnamefont{S.}~\bibnamefont{Ryu}} \bibnamefont{and}
  \bibinfo{author}{\bibfnamefont{Y.}~\bibnamefont{Hatsugai}},
  \bibinfo{journal}{Phys.\ Rev.\ Lett.} \textbf{\bibinfo{volume}{89}},
  \bibinfo{pages}{077002} (\bibinfo{year}{2002}).

\bibitem{commentcircle}  
$\Delta $ delineates a circle centered at 
$C_0=1 + e^{ik_2}$ with a radius 
$r=\sqrt{1+2(t'/t)\cos k_2+{(t'/t)}^2}$
in the complex $\Delta$-plane 
when $k_1$ is varied from $0$ to $2\pi$.  
By examining $C_0$ and $r$, we can show that we have a 
Dirac (linear) dispersion when 
the circle goes through the origin, which occurs for $-3\le t'/t \le
		   1$. 
\bibitem[{\citenamefont{Hofstadter}(1976)}]{Hof76}
\bibinfo{author}{\bibfnamefont{D. R. }~\bibnamefont{Hofstadter}},
  \bibinfo{journal}{Phys. \ Rev.\ B} \textbf{\bibinfo{volume}{14}},
  \bibinfo{pages}{2239} (\bibinfo{year}{1976}).


\bibitem[{\citenamefont{Ryu and Hatsugai}(2003)}]{Ryu03cb}
\bibinfo{author}{\bibfnamefont{S.}~\bibnamefont{Ryu}} \bibnamefont{and}
  \bibinfo{author}{\bibfnamefont{Y.}~\bibnamefont{Hatsugai}},
  \bibinfo{journal}{Physica C} \textbf{\bibinfo{volume}{388-389}},
  \bibinfo{pages}{90} (\bibinfo{year}{2003}).

\bibitem[{\citenamefont{Sheng Sheng Weng}(2003)}]{SSW06}
\bibinfo{author}{\bibfnamefont{D. N. }~\bibnamefont{Sheng}}, 
\bibinfo{author}{\bibfnamefont{L. }~\bibnamefont{Sheng}}, \bibnamefont{and}
  \bibinfo{author}{\bibfnamefont{Z. Y.}~\bibnamefont{Weng}},
  \bibinfo{journal}{cond-mat/0602190}. \textbf{\bibinfo{volume}{}}
  \bibinfo{pages}{}



\bibitem{kh}
Topological QHE numbers when the transfer energies along three 
directions in the honeycomb lattice are varied have recently been 
discussed by
Y. Hasegawa, R. Konno, H. Nakano and M. Kohmoto, cond-mat/0604433,
and 
Y. Hasegawa and M. Kohmoto, cond-mat/0603345.


\bibitem[{\citenamefont{Rammal}(1985)}]{ram}
\bibinfo{author}{\bibfnamefont{R.}~\bibnamefont{Rammal}}, \bibinfo{journal}{J.
  Physique} \textbf{\bibinfo{volume}{46}}, \bibinfo{pages}{1345}
  (\bibinfo{year}{1985}).

\bibitem[{\citenamefont{Thouless et~al.}(1982)\citenamefont{Thouless, Kohmoto,
  Nightingale, and den Nijs}}]{Thouless82}
\bibinfo{author}{\bibfnamefont{D.~J.} \bibnamefont{Thouless}},
  \bibinfo{author}{\bibfnamefont{M.}~\bibnamefont{Kohmoto}},
  \bibinfo{author}{\bibfnamefont{P.}~\bibnamefont{Nightingale}},
  \bibnamefont{and} \bibinfo{author}{\bibfnamefont{M.}~\bibnamefont{den Nijs}},
  \bibinfo{journal}{Phys.\ Rev.\ Lett.} \textbf{\bibinfo{volume}{49}},
  \bibinfo{pages}{405} (\bibinfo{year}{1982}).

\bibitem[{\citenamefont{Kohmoto}(1985)}]{Kohmoto85}
\bibinfo{author}{\bibfnamefont{M.}~\bibnamefont{Kohmoto}},
  \bibinfo{journal}{Ann.\ Phys.\ (N. Y. )} \textbf{\bibinfo{volume}{160}},
  \bibinfo{pages}{355} (\bibinfo{year}{1985}).

\bibitem[{\citenamefont{Aoki and Ando}(1981)}]{Aoki81}
\bibinfo{author}{\bibfnamefont{H.}~\bibnamefont{Aoki}} \bibnamefont{and}
  \bibinfo{author}{\bibfnamefont{T.}~\bibnamefont{Ando}},
  \bibinfo{journal}{Solid\ State \ Comm.} \textbf{\bibinfo{volume}{38}},
  \bibinfo{pages}{1079} (\bibinfo{year}{1981}).

\bibitem[{\citenamefont{Fukui et~al.}(2005)\citenamefont{Fukui, Hatsugai, and
  Suzuki}}]{Fukui05}
\bibinfo{author}{\bibfnamefont{T.}~\bibnamefont{Fukui}},
  \bibinfo{author}{\bibfnamefont{Y.}~\bibnamefont{Hatsugai}}, \bibnamefont{and}
  \bibinfo{author}{\bibfnamefont{H.}~\bibnamefont{Suzuki}},
  \bibinfo{journal}{J.\ Phys. \ Soc.\ Jpn.} \textbf{\bibinfo{volume}{74}},
  \bibinfo{pages}{1674} (\bibinfo{year}{2005}).

\bibitem[{\citenamefont{Hatsugai et~al.}(2005)\citenamefont{Hatsugai, Fukui,
  and Suzuki}}]{Hat05ep2ds}
\bibinfo{author}{\bibfnamefont{Y.}~\bibnamefont{Hatsugai}},
  \bibinfo{author}{\bibfnamefont{T.}~\bibnamefont{Fukui}}, \bibnamefont{and}
  \bibinfo{author}{\bibfnamefont{H.}~\bibnamefont{Suzuki}},
  \bibinfo{journal}{cond-mat/0507466}  (\bibinfo{year}{2005}).

\bibitem[{\citenamefont{Luscher}(1982)}]{Lus82}
\bibinfo{author}{\bibfnamefont{M.}~\bibnamefont{L\"uscher}},
  \bibinfo{journal}{Commun.\ Math.\ Phys.} \textbf{\bibinfo{volume}{85}},
  \bibinfo{pages}{39} (\bibinfo{year}{1982}).

\bibitem[{\citenamefont{Phillips}(1985)}]{Phi85}
\bibinfo{author}{\bibfnamefont{A. }~\bibnamefont{Phillips}},
  \bibinfo{journal}{Ann.\ Phys.\ (N. Y.)} \textbf{\bibinfo{volume}{161}},
  \bibinfo{pages}{399} (\bibinfo{year}{1985}).

\bibitem[{\citenamefont{Phillips}(1986)}]{PhiSto86}
\bibinfo{author}{\bibfnamefont{A. }~\bibnamefont{Phillips}}  \bibnamefont{and}
  \bibinfo{author}{\bibfnamefont{D. }~\bibnamefont{Stone}},
  \bibinfo{journal}{Commun.\ Math.\ Phys.} \textbf{\bibinfo{volume}{103}},
  \bibinfo{pages}{599} (\bibinfo{year}{1986}).

\bibitem[{\citenamefont{Phillips}(1990)}]{PhiSto90}
\bibinfo{author}{\bibfnamefont{A. }~\bibnamefont{Phillips}}  \bibnamefont{and}
  \bibinfo{author}{\bibfnamefont{D. }~\bibnamefont{Stone}},
  \bibinfo{journal}{Commun.\ Math.\ Phys.} \textbf{\bibinfo{volume}{131}},
  \bibinfo{pages}{255} (\bibinfo{year}{1990}).

\bibitem[{\citenamefont{L\"uscher}(1999)}]{Lus99}
\bibinfo{author}{\bibfnamefont{M. }~\bibnamefont{L\"uscher}},
  \bibinfo{journal}{Nucl.\ Phys.\ B} \textbf{\bibinfo{volume}{538}},
  \bibinfo{pages}{515} (\bibinfo{year}{1999}).

\bibitem[{\citenamefont{Fujiwara}(2001)\citenamefont{Fujiwara,
		   H. Suzuki, and K. Wu}}]{FSW01}
\bibinfo{author}{\bibfnamefont{T. }~\bibnamefont{Fujiwara}},
  \bibinfo{author}{\bibfnamefont{H.}~\bibnamefont{Suzui}}, \bibnamefont{and}
  \bibinfo{author}{\bibfnamefont{K.}~\bibnamefont{Wu}},
  \bibinfo{journal}{Prog.\ Theor. \ Phys.\ } \textbf{\bibinfo{volume}{105}},
  \bibinfo{pages}{789} (\bibinfo{year}{2005}).

\bibitem[{\citenamefont{Hatsugai}(1997)}]{yh-qhe-review}
\bibinfo{author}{\bibfnamefont{Y.}~\bibnamefont{Hatsugai}},
  \bibinfo{journal}{J.\ Phys. \ C,\ Condens. \ Matter}
  \textbf{\bibinfo{volume}{9}}, \bibinfo{pages}{2507} (\bibinfo{year}{1997}).

\bibitem{kusakabe}
K. Kusakabe and Y. Takagi, Mol. Cryst. Liq. Cryst. 387, 7 (2002).
 
\bibitem{aoki92} H. Aoki, Surf. Sci. 263, 137 (1992).  


\end{thebibliography}

\end{document}